\documentclass[a4paper,fleqn,usenatbib]{mnras}

\usepackage{newtxtext,newtxmath}

\usepackage[T1]{fontenc}
\usepackage{ae,aecompl}

\usepackage{graphicx}	
\usepackage{amsmath}	

\title[Period Changes for 14 Novae]{Orbital Period Changes for Fourteen Novae and the Critical Failures of the Predictions of Standard Theories, the Hibernation Model, and the Magnetic Braking Model}

\author[B. E. Schaefer]{
Bradley E. Schaefer$^{1}$\thanks{E-mail: schaefer@lsu.edu},
\\
$^{1}$Department of Physics and Astronomy, Louisiana State University, Baton Rouge, Louisiana, 70820, USA\\
}

\pubyear{2021}

\begin{document}
\label{firstpage}
\pagerange{\pageref{firstpage}--\pageref{lastpage}}
\maketitle

\begin{abstract}

The evolution of novae and Cataclysmic Variables (CVs) is driven by changes in the binary orbital periods.  In a direct and critical test for various evolution models and their physical mechanisms, I measure the sudden changes in the period ($\Delta P$) across 14 nova eruptions and I measure the steady period change during quiescence ($\dot{P}$) for 20 inter-eruption intervals.  The standard theory for $\Delta P$ is dominated by the mechanism of mass loss, and this fails completely for the five novae with {\it negative} values, and it fails to permit the $\Delta P$ for U Sco eruptions to change by one order-of-magnitude from eruption-to-eruption.  The Hibernation Model of evolution is refuted because all the $\Delta P$ measures are orders of magnitude too small to cause any significant drop in accretion luminosity, and indeed, near half of the nova have {\it negative} $\Delta P$ as the opposite of the required mechanism for any hibernation state.  As for the Magnetic Braking Model, this fails by many orders-of-magnitude in its predictions of the required $\dot{P}$ for 9-out-of-13 novae.  The observed $\dot{P}$ values scatter, both positively and negatively, over a range of $\pm$10$^{-9}$, while the predicted values are from $-$10$^{-13}$ to $-$10$^{-11}$.  This huge scatter is not possible with standard theory, and there must be some currently-unknown mechanism to be added in, with this new mechanism 100--10000$\times$ larger in effect than the current theory allows.  In all, these failed predictions demonstrate that nova systems must have unknown physical mechanisms for both $\Delta P$ and $\dot{P}$ that dominate over all other effects.
 
\end{abstract}




\section{INTRODUCTION}

Cataclysmic variables (CVs), including classical novae (CNe) and recurrent novae (RNe), now have the foremost science issue as to their evolution.  CV evolution is driven and measured by the orbital period, $P$, and by the long-term changes of $P$.  The period changes are either slow and steady, measured as the time derivative $\dot{P}$, or are the sudden period change across a nova eruption, measured as $\Delta P$.  CV evolution can only be understood by measuring the period changes of many CVs and matching the observed results to a physical theory.

The theory of CV evolution consists of physical models of the period changes and its consequences.  The long-time consensus model is called the `Magnetic Braking Model' (MBM, Knigge, Baraffe, \& Patterson 2011).  This model provides an empirical power-law prescription for the loss of the binary angular momentum, leading to a derived evolutionary track of $\dot{P}$ as a function of $P$.  Our community accepts the MBM as a consensus because it provides an explanation of the `Period Gap' (where few CVs have $P$ in the roughly 2--3 hour range), it explains the minimum orbital $P$ for CVs, and it describes the inevitable  overall evolution from long-period to short-period.  

Further, superposed on the MBM evolutionary track, the controversial and alluring `Hibernation Model' (HibM) adds a middle-term cycle (Shara et al. 1986).   In the Hibernation Model, each CV cycles through a nova-like state, through a nova eruption, then a falling accretion rate moves the system through a dwarf nova phase into a detached hibernation phase, only to have the binary come back into contact through slow angular momentum loss mechanisms, return to the nova-like state to start the cycle over again.  Critically, the driving force for this cycle is a large and positive $\Delta P$ from each nova event, with this kick to the system resulting in an increased binary separation and the falling accretion rate.  In the original Hibernation Model, all novae and CVs are supposed to cycle through the detached state of deep hibernation.  A revised Hibernation Model (Hillman et al. 2020) now backtracks and claims that hibernation occurs only for the shortest-period CVs.  In all cases, Hibernation is driven by the CV period changes, and the predictions should be tested against many observed values of $\Delta P$.

On the observational side, we have had few reliable measures of $\dot{P}$ or $\Delta P$.  These measures all come from the timing of many eclipse epochs (or times of minimum light) that serve as markers of the orbital position of the companion star, with the traditional analysis based on the $O-C$ diagram.  The problem for $\dot{P}$ is that many-decades of orbital period timings are required to get a measure that dominates over the various sources of noise.  The problem for $\Delta P$ is that we must somehow get decades of eclipse timings from {\it before} the star was known to be a nova.  Both of these problems can only be solved by using archival data, as the only way to get long-enough of a period record to obtain reliable period changes.  The first reliable measure of an evolutionary period change in any nova was for BT Mon (Nova Mon 1939), for which Schaefer \& Patterson (1983) used the archival plates from 1905--1939 at Harvard College Observatory to get many pre-eruption eclipse times and hence $\Delta P$.  BT Mon had a period increase of $+$40 parts-per-million (ppm), with this one measure serving as a substantial part of the motivation for the original Hibernation Model.  Before and after this BT Mon result, a variety of papers have appeared showing $O-C$ curves for novae, but unfortunately these all have problems making for no reliable measure of the evolutionary period changes.  The result is that, until recently, our field has had only one reliable measure of an evolutionary period change in a nova (for BT Mon), and such is not adequate to provide the fundamental test of the cornerstone models of CV evolution.

Starting in 1989, I have been pursuing a career-long program of measuring eclipse and minima times in RNe and later CNe for purposes of determining $\Delta P$.  Originally, I made eclipse timings of RNe because I could then know in advance which stars would have future eruptions.   My original motivation was to measure $\Delta P$, derive the mass of the ejected shell, test to see whether the RN was ejecting more mass than was being accreted on to the white dwarf, all as a test of the seductive idea that RNe are the progenitors of Type {\rm I}a supernovae.  Around the year 2013, I realized that an intensive search of archival plates worldwide could reveal old eclipse times and pre-eruption orbital periods for half-a-dozen old novae so as to yield new measures of $\Delta P$.  As a side-product of this work, I realized that I had to measure the $\dot{P}$ for each nova, and such long-term values can serve as a fundamental test of the cornerstone Magnetic Braking Model.  Further, I started work on novae for which I only could measure $\dot{P}$.  This long and tedious work has resulted in $\dot{P}$ or $\Delta P$ measures for 6 CNe (Schaefer 2020b) and 4 RNe (Schaefer 2011; 2022a; 2023; Schaefer et al. 2013).

CV evolution is driven by the period changes $\dot{P}$ and $\Delta P$.  Our community has long had well-developed and compelling models based on predicting these period changes.  But the primary driver of these models (i.e., the period changes) has not been seriously tested.  That is, other than my few recent papers presenting long-term measures of $\dot{P}$ and $\Delta P$, the predicted and required period changes of novae have not been tested.  So the front-line of CV research is now concerning CV evolution, with this needing real testing of the model predictions versus measured period changes.


\section{CI Aql}

\begin{table}
	\centering
	\caption{CI Aql Eclipse Times.  Full table of 125 lines in Supplementary Material.}
	\begin{tabular}{lllrr}
		\hline
		Eclipse Time (HJD) & Source   &   Year   &   $N$   &   $O-C$ \\
		\hline
2424712.7343	$\pm$	0.0300	&	HCO	&	1926.539	&	-43593	&	-0.1335	\\
2428012.3059	$\pm$	0.0300	&	HCO	&	1935.572	&	-38257	&	-0.1335	\\
2448420.8002	$\pm$	0.0080	&	RoboScope	&	1991.448	&	-5253	&	-0.0095	\\
2448469.6472	$\pm$	0.0065	&	RoboScope	&	1991.582	&	-5174	&	-0.0130	\\
2448495.6270	$\pm$	0.0080	&	RoboScope	&	1991.653	&	-5132	&	-0.0044	\\
...   &  ...   &   ...   &   ...   &   ...   \\
2455674.7973	$\pm$	0.0006	&	CTIO 1.3-m	&	2011.308	&	6478	&	0.0004	\\
2455682.8335	$\pm$	0.0003	&	CTIO 1.3-m	&	2011.330	&	6491	&	-0.0021	\\
2455687.7820	$\pm$	0.0027	&	CTIO 1.3-m	&	2011.344	&	6499	&	-0.0004	\\
2455695.8224	$\pm$	0.0005	&	CTIO 1.3-m	&	2011.366	&	6512	&	0.0013	\\
2455700.7656	$\pm$	0.0005	&	CTIO 1.3-m	&	2011.379	&	6520	&	-0.0024	\\
2455705.7146	$\pm$	0.0004	&	CTIO 1.3-m	&	2011.393	&	6528	&	-0.0003	\\
2455708.8040	$\pm$	0.0005	&	CTIO 1.3-m	&	2011.401	&	6533	&	-0.0027	\\
2455713.7520	$\pm$	0.0027	&	CTIO 1.3-m	&	2011.415	&	6541	&	-0.0016	\\
2455783.6292	$\pm$	0.0020	&	CTIO 1.3-m	&	2011.606	&	6654	&	0.0009	\\
2456066.8351	$\pm$	0.0003	&	CTIO 1.3-m	&	2012.381	&	7112	&	-0.0023	\\
2456071.7831	$\pm$	0.0004	&	CTIO 1.3-m	&	2012.395	&	7120	&	-0.0012	\\
2456149.6950	$\pm$	0.0003	&	CTIO 1.3-m	&	2012.608	&	7246	&	-0.0028	\\
2456154.6409	$\pm$	0.0004	&	CTIO 1.3-m	&	2012.622	&	7254	&	-0.0037	\\
2456159.5891	$\pm$	0.0003	&	CTIO 1.3-m	&	2012.635	&	7262	&	-0.0024	\\
2456167.0090	$\pm$	0.0001	&	SSO 2.3-m	&	2012.656	&	7274	&	-0.0029	\\
2456416.8288	$\pm$	0.0006	&	CTIO 1.3-m	&	2013.340	&	7678	&	-0.0007	\\
2456442.7980	$\pm$	0.0003	&	CTIO 1.3-m	&	2013.411	&	7720	&	-0.0026	\\
2456447.7442	$\pm$	0.0004	&	CTIO 1.3-m	&	2013.424	&	7728	&	-0.0033	\\
2456452.6937	$\pm$	0.0004	&	CTIO 1.3-m	&	2013.438	&	7736	&	-0.0007	\\
2456455.7849	$\pm$	0.0004	&	CTIO 1.3-m	&	2013.446	&	7741	&	-0.0013	\\
2456460.7317	$\pm$	0.0004	&	CTIO 1.3-m	&	2013.460	&	7749	&	-0.0014	\\
2458720.8311	$\pm$	0.0008	&	ZTF	&	2019.648	&	11404	&	-0.0096	\\
2459018.8807	$\pm$	0.0007	&	ZTF	&	2020.464	&	11886	&	-0.0098	\\
2459064.6354	$\pm$	0.0021	&	AAVSO	&	2020.589	&	11960	&	-0.0138	\\
2459066.4908	$\pm$	0.0016	&	AAVSO	&	2020.594	&	11963	&	-0.0135	\\
2459072.6748	$\pm$	0.0024	&	AAVSO	&	2020.611	&	11973	&	-0.0131	\\
2459077.6230	$\pm$	0.0017	&	AAVSO	&	2020.625	&	11981	&	-0.0117	\\
2459089.9930	$\pm$	0.0005	&	AAVSO	&	2020.658	&	12001	&	-0.0089	\\
2459095.5538	$\pm$	0.0018	&	AAVSO	&	2020.674	&	12010	&	-0.0134	\\
2459098.0309	$\pm$	0.0003	&	AAVSO	&	2020.680	&	12014	&	-0.0097	\\
2459100.5033	$\pm$	0.0016	&	AAVSO	&	2020.687	&	12018	&	-0.0108	\\
2459347.8443	$\pm$	0.0008	&	AAVSO	&	2021.364	&	12418	&	-0.0140	\\
2459355.8851	$\pm$	0.0012	&	AAVSO	&	2021.386	&	12431	&	-0.0119	\\
2459357.7395	$\pm$	0.0013	&	AAVSO	&	2021.391	&	12434	&	-0.0126	\\
2459365.7803	$\pm$	0.0019	&	AAVSO	&	2021.413	&	12447	&	-0.0105	\\
2459367.6356	$\pm$	0.0014	&	AAVSO	&	2021.419	&	12450	&	-0.0103	\\
2459393.6044	$\pm$	0.0010	&	AAVSO	&	2021.490	&	12492	&	-0.0126	\\
2459396.6964	$\pm$	0.0009	&	AAVSO	&	2021.498	&	12497	&	-0.0124	\\
2459407.8262	$\pm$	0.0005	&	ZTF	&	2021.529	&	12515	&	-0.0130	\\
2459422.0455	$\pm$	0.0004	&	AAVSO	&	2021.568	&	12538	&	-0.0161	\\
2459422.6720	$\pm$	0.0012	&	AAVSO	&	2021.569	&	12539	&	-0.0079	\\
2459424.5233	$\pm$	0.0012	&	AAVSO	&	2021.574	&	12542	&	-0.0117	\\
2459739.8841	$\pm$	0.0009	&	ZTF	&	2022.438	&	13052	&	-0.0148	\\
2459805.4307	$\pm$	0.0008	&	AAVSO	&	2022.617	&	13158	&	-0.0144	\\
2459858.6073	$\pm$	0.0005	&	AAVSO	&	2022.763	&	13244	&	-0.0168	\\
		\hline
	\end{tabular}	
\end{table}

\begin{figure}
	\includegraphics[width=1.0\columnwidth]{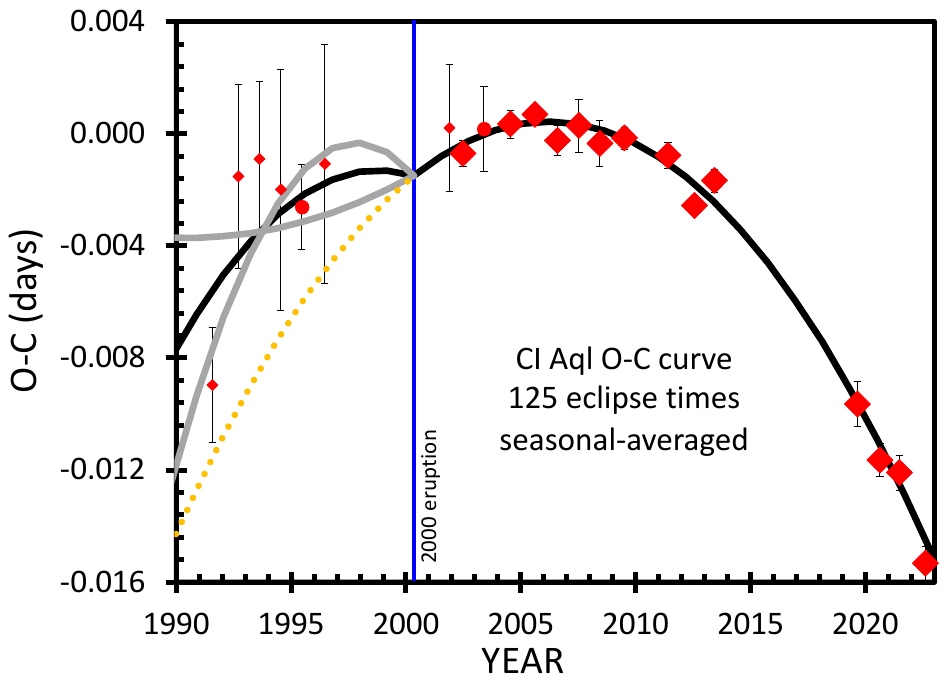}
    \caption{$O-C$ curve for CI Aql.  The fiducial linear ephemeris uses an epoch of HJD 2451669.0575 and a period of 0.61836051 d.  The red diamonds are the seasonal averages from 125 eclipse times.  The best-fitting model is shown as a thick black curve, being a broken parabola with the break at the time of the 2000 eruption (shown as a thin blue vertical line).  The post-eruption $O-C$ curve has a well-defined curvature, with $\dot{P}_{post}$=($-$5.16$\pm$0.45)$\times$10$^{-10}$, with the negative sign meaning that the orbital period suffers a steady {\it decrease}.  The post-eruption model is extended to before 2000 as shown by the dotted orange line.  The pre-eruption curve is poorly defined, with the one-sigma range on $\dot{P}_{pre}$ displayed by the two thick gray parabolas.  The pre-eruption curvature is consistent with the post-eruption curvature and with positive curvature.  The sudden change of the orbital period across the 2000 eruption, $\Delta P$, can be seen as the sharp kink in the $O-C$ curve.  The best-fitting $\Delta P$ ($+$0.00000154 d) is positive, indicating that the eruption resulted in a $+$2.5 ppm {\it increase} in the period.  The one-sigma range is from $+$0.6 to $+$4.4 ppm, so the existence of a non-zero $\Delta P$ is not highly significant.}
\end{figure}

CI Aql is a recurrent nova with known eruptions in 1917, 1941, and 2000, reaching a peak of $V$=9.0 (rising from a quiescent level near $V$=16.7), and with a relatively slow fade from peak by three magnitudes ($t_3$) in 32 days (Schaefer 2010).  With an average recurrence time-scale of 24 years, CI Aql is expected to have another eruption any month now.  {\it Before} the second discovered eruption in 2000, with their Roboscope automated observatory, Mennickent \& Honeycutt (1995) found the eclipses for an orbital period of 0.618355 d.  

With the eruption of 2000, the Mennickent \& Honeycutt light curve serves to provide a measure of the {\it pre}-eruption orbital period, $P_{\rm pre}$.  Soon after the 2000 eruption, I began a long series of eclipse timings with the telescopes at the McDonald and Cerro Tololo observatories, with the purpose of measuring the post-eruption period, $P_{\rm post}$ (Schaefer 2011).  The result was a poorly defined $P_{\rm pre}$, a moderately good $P_{post}$, and a period change across the eruption ($\Delta P$=$P_{\rm post}$-$P_{\rm pre}$) that was consistent with zero.  I have continued my series of eclipse timings with the Cerro Tololo 1.3-m and the Siding Springs 2.5-m telescopes.  Further, I have collected light curves from the American Association of Variable Star Observers (AAVSO) International Database\footnote{\url{https://www.aavso.org/data-download}} (AID) and from the Zwicky Transient Factory\footnote{\url{https://irsa.ipac.caltech.edu/cgi-bin/Gator/nph-scan?projshort=ZTF}} (ZTF), with these returning eclipse times from 2019--2022.  The print version of Table 1 displays eclipse times, as heliocentric Julian Dates (HJD), including all of my new eclipse times.  The remainder are listed in Schaefer (2011).  All 125 CI Aql eclipse times are in the Supplementary Material  version of Table 1.  

The $O-C$ value is the time difference (in days) between the observed eclipse time and the calculated eclipse time for some fiducial linear ephemeris.  For CI Aql, I choose the ephemeris with an epoch of HJD 2451669.0575 and a period of 0.61836051 d.  The observed $O-C$ curve in Fig. 1 has each red diamond giving a seasonal average.  

The scatter within a single observing season is always greatly larger than the reported measurement errors (as in Table 1), demonstrating that there is some additional source of timing error superposed on top of the measurement uncertainty.  This additional error source arises from the inevitable jitter in the time of photometric minimum around the time of conjunction caused by the ubiquitous and ordinary flickering in the light curve skewing each and every eclipse light curve (Schaefer 2021).  The RMS scatter of eclipse times around any smooth $O-C$ curve has a value of 0.0014 d above that expected by the reported measurement errors.  This means that the flickering in the CI Aql light curve contributes a source of uncertainty of 0.0014 d that must be added in quadrature to the measurement error so as to get the total error for each eclipse time.  

The post-eruption $O-C$ curve can be accurately fitted to a parabola, with the parabola describing the case of a constant $\dot{P}$.  The parabola model assumes that the eclipse times are given by the equation
\begin{equation}
T = E_0 + P N + 0.5 P \dot{P} N^2,
\end{equation}
with all the terms having units of `day'.  The $E_0$ is an epoch in HJD for one eclipse time that serves as the `zero epoch'.  The $N$ value is an integer that identifies each eclipse.  The $\dot{P}$ value is the steady period change, with units of days/day or dimensionless.  

The post-eruption $O-C$ is shown in Fig. 1, along with the best-fitting parabolic model.  The reduced $\chi^2$ is close to unity, showing both that the eclipse timing error bars are accurate and that CI Aql has a steady $\dot{P}$.  The best-fitting $\dot{P}$ is ($-$5.16$\pm$0.45)$\times$10$^{-10}$, where the negative sign shows that the CI Aql period is {\it decreasing} over time.  The orbital period just after eruption is $P_{\rm post}$=0.61836162$\pm$0.00000022 d.  The eclipse time for $N$=0 in the year 2000 is HJD 2451669.0560$\pm$0.0007.  

This improved $E_0$ serves as the last eclipse time for the pre-eruption time interval.  Now, with an accurate measure, $E_0$ provides a strong anchor for using the light curve of Mennickent \& Honeycutt to get $P_{\rm pre}$ and $\dot{P}_{\rm pre}$.  Previously, I had used the eclipse times (as in Table 1) to estimate the pre-eruption period.  But now, I have used all 285 magnitudes from 1991--1996 in a $\chi^2$ fit to a template light curve shape.  My template is from Schaefer (2011).  This procedure includes the information from the secondary eclipses and the ellipsoidal effects in addition to the primary eclipse shape, so as to produce better values for $P_{\rm pre}$ and $\dot{P}_{\rm pre}$.  With this, the best period immediately before the 2000 eruption is $P_{\rm pre}$=0.61836008$\pm$0.00000022 d.   The best $\dot{P}_{\rm pre}$ is ($-$4.7$\pm$6.0)$\times$10$^{-10}$.  Pointedly, the one-sigma confidence interval includes zero and positive values.  The pre-eruption models with one-sigma extremes in $\dot{P}_{\rm pre}$ are shown as thick gray curves (each a parabola segment) in Fig. 1.

I have made substantial improvements in $E_0$ and in fitting the pre-eruption light curve, resulting in a $P_{\rm pre}$ value with much better accuracy than in Schaefer (2011).  With this, $\Delta P$ is $+$0.00000154 d, which is a fractional change of $+$2.5 parts-per-million (ppm).  The one-sigma range for $\dot{P}_{\rm pre}$ has $\Delta P$/$P$ changes from $+$0.6 to $+$4.4 ppm, so the positive sign for $\Delta P$ is not highly significant.  Nevertheless, to get a negative $\Delta P$, the $\dot{P}_{\rm pre}$ would have substantial concave-up curvature, which seems unlikely given the well-measured $\dot{P}_{\rm post}$.  


\section{T AUR}

T Aur (Nova Aurigae 1891) was discovered with the unaided eye at a magnitude near 4.5 by the Scottish amateur T. D. Anderson, with this kickstarting the discovery mode of amateurs stepping outside nightly and making deep dome searches, with this mode dominating until the end of World War {\rm II}.  With Anderson's quick notice, T Aur provided the first photographic spectrum of a nova, allowing detailed and quantitative measures.  T Aur was the first nova with a `dust dip', where the nova light suffered a fast and deep fade due to dust formation inside the shell of ejecta dimming the light on the star in the middle.  T Aur was the third old nova discovered to be a close binary system, with this case allowing the generalization to all novae (Walker 1962).  Walker's discovery that T Aur is an eclipsing binary also provides the earliest eclipse times for use in my $O-C$ curve.


\begin{table}
	\centering
	\caption{T Aur Eclipse Times.  Full Table in Supplementary Material.}
	\begin{tabular}{lllrr}
		\hline
		Eclipse Time (HJD) & Source   &   Year   &   $N$   &   $O-C$ \\
		\hline
2434797.6760	$\pm$	0.0010	&	Walker	&	1954.149	&	-13780	&	-0.0041	\\
2437614.0110	$\pm$	0.0010	&	Walker	&	1961.860	&	0	&	-0.0012	\\
...   &  ...   &   ...   &   ...   &   ...   \\
2454468.4740	$\pm$	0.0006	&	AAVSO	&	2008.005	&	82467	&	0.0019	\\
2454468.4747	$\pm$	0.0007	&	AAVSO	&	2008.005	&	82467	&	0.0026	\\
2455851.5037	$\pm$	0.0006	&	AAVSO	&	2011.791	&	89234	&	0.0041	\\
2457068.5761	$\pm$	0.0008	&	AAVSO	&	2015.123	&	95189	&	0.0041	\\
2457069.3915	$\pm$	0.0003	&	AAVSO	&	2015.126	&	95193	&	0.0020	\\
2457074.5016	$\pm$	0.0007	&	AAVSO	&	2015.140	&	95218	&	0.0027	\\
2457081.6531	$\pm$	0.0005	&	AAVSO	&	2015.159	&	95253	&	0.0008	\\
2458013.6173	$\pm$	0.0004	&	AAVSO	&	2017.711	&	99813	&	0.0004	\\
2458020.5687	$\pm$	0.0004	&	AAVSO	&	2017.730	&	99847	&	0.0029	\\
2458041.4149	$\pm$	0.0010	&	Mazanec	&	2017.787	&	99949	&	0.0025	\\
2458041.4154	$\pm$	0.0010	&	Mazanec	&	2017.787	&	99949	&	0.0030	\\
2458041.8230	$\pm$	0.0006	&	AAVSO	&	2017.788	&	99951	&	0.0018	\\
2458042.8445	$\pm$	0.0005	&	AAVSO	&	2017.791	&	99956	&	0.0014	\\
2458043.8674	$\pm$	0.0005	&	AAVSO	&	2017.794	&	99961	&	0.0024	\\
2458045.9109	$\pm$	0.0004	&	AAVSO	&	2017.799	&	99971	&	0.0021	\\
2458052.8582	$\pm$	0.0004	&	AAVSO	&	2017.818	&	100005	&	0.0006	\\
2458058.7859	$\pm$	0.0005	&	AAVSO	&	2017.834	&	100034	&	0.0013	\\
2458060.8294	$\pm$	0.0005	&	AAVSO	&	2017.840	&	100044	&	0.0011	\\
2458071.8651	$\pm$	0.0005	&	AAVSO	&	2017.870	&	100098	&	0.0003	\\
2458072.6832	$\pm$	0.0005	&	AAVSO	&	2017.872	&	100102	&	0.0009	\\
2458482.6685	$\pm$	0.0006	&	AAVSO	&	2018.995	&	102108	&	0.0035	\\
2458489.6182	$\pm$	0.0006	&	AAVSO	&	2019.014	&	102142	&	0.0043	\\
2458493.7043	$\pm$	0.0004	&	AAVSO	&	2019.025	&	102162	&	0.0028	\\
2458493.7047	$\pm$	0.0020	&	AAVSO	&	2019.025	&	102162	&	0.0032	\\
2458495.7490	$\pm$	0.0005	&	AAVSO	&	2019.031	&	102172	&	0.0037	\\
2458505.5550	$\pm$	0.0006	&	AAVSO	&	2019.058	&	102220	&	-0.0004	\\
2458523.5436	$\pm$	0.0005	&	AAVSO	&	2019.107	&	102308	&	0.0029	\\
2458770.8436	$\pm$	0.0005	&	AAVSO	&	2019.784	&	103518	&	0.0053	\\
2458785.7582	$\pm$	0.0004	&	AAVSO	&	2019.825	&	103591	&	0.0003	\\
2458785.9649	$\pm$	0.0003	&	AAVSO	&	2019.825	&	103592	&	0.0026	\\
2458788.0070	$\pm$	0.0003	&	AAVSO	&	2019.831	&	103602	&	0.0009	\\
2458791.8913	$\pm$	0.0003	&	AAVSO	&	2019.842	&	103621	&	0.0021	\\
2458793.9351	$\pm$	0.0004	&	AAVSO	&	2019.847	&	103631	&	0.0020	\\
2458800.8863	$\pm$	0.0005	&	AAVSO	&	2019.866	&	103665	&	0.0044	\\
2458804.9705	$\pm$	0.0004	&	AAVSO	&	2019.877	&	103685	&	0.0010	\\
2458808.8528	$\pm$	0.0005	&	AAVSO	&	2019.888	&	103704	&	0.0001	\\
2458809.8782	$\pm$	0.0009	&	AAVSO	&	2019.891	&	103709	&	0.0037	\\
2458810.6978	$\pm$	0.0006	&	AAVSO	&	2019.893	&	103713	&	0.0057	\\
2458813.7607	$\pm$	0.0006	&	AAVSO	&	2019.901	&	103728	&	0.0029	\\
2458817.8489	$\pm$	0.0003	&	AAVSO	&	2019.913	&	103748	&	0.0036	\\
2458824.7972	$\pm$	0.0004	&	AAVSO	&	2019.932	&	103782	&	0.0030	\\
2458827.8620	$\pm$	0.0005	&	AAVSO	&	2019.940	&	103797	&	0.0021	\\
2458829.08846	$\pm$	0.00012	&	TESS 19	&	2019.944	&	103803	&	0.0023	\\
2458836.8619	$\pm$	0.0007	&	AAVSO	&	2019.965	&	103841	&	0.0094	\\
2458848.3013	$\pm$	0.0004	&	AAVSO	&	2019.996	&	103897	&	0.0037	\\
2458849.7358	$\pm$	0.0006	&	AAVSO	&	2020.000	&	103904	&	0.0075	\\
2458850.5504	$\pm$	0.0005	&	AAVSO	&	2020.002	&	103908	&	0.0045	\\
2458850.7517	$\pm$	0.0005	&	AAVSO	&	2020.003	&	103909	&	0.0014	\\
2458854.6364	$\pm$	0.0004	&	AAVSO	&	2020.013	&	103928	&	0.0030	\\
2458855.6601	$\pm$	0.0005	&	AAVSO	&	2020.016	&	103933	&	0.0048	\\
2458861.3795	$\pm$	0.0004	&	AAVSO	&	2020.032	&	103961	&	0.0017	\\
2458869.5553	$\pm$	0.0004	&	AAVSO	&	2020.054	&	104001	&	0.0023	\\
2458888.5599	$\pm$	0.0005	&	AAVSO	&	2020.106	&	104094	&	-0.0003	\\
2459210.6652	$\pm$	0.0005	&	AAVSO	&	2020.988	&	105670	&	0.0049	\\
2459226.6048	$\pm$	0.0004	&	AAVSO	&	2021.032	&	105748	&	0.0031	\\
2459232.5297	$\pm$	0.0003	&	AAVSO	&	2021.048	&	105777	&	0.0010	\\
2459234.5754	$\pm$	0.0004	&	AAVSO	&	2021.054	&	105787	&	0.0029	\\
2459234.7777	$\pm$	0.0004	&	AAVSO	&	2021.054	&	105788	&	0.0008	\\
2459486.16385	$\pm$	0.00012	&	TESS 43	&	2021.743	&	107018	&	0.0017	\\
2459512.12041	$\pm$	0.00013	&	TESS 44	&	2021.814	&	107145	&	0.0022	\\
2459538.07588	$\pm$	0.00011	&	TESS 45	&	2021.885	&	107272	&	0.0017	\\
		\hline
	\end{tabular}	
\end{table}

I have compiled a list of eclipse times (see Table 2).  These are all determined with parabola fits to the light curve for the times around mid-eclipse.  For eclipse times before 2009 that have appeared in the literature, Dai \& Qian (2010) provide a convenient collection.  Further, I have collected light curves for 55 eclipses from 2008 to 2021 from the AAVSO AID.  These times are averaged together on a yearly basis to give seasonal averages for the $O-C$ curve.  Further, I have used the light curves from the {\it TESS} spacecraft\footnote{\url{https://mast.stsci.edu/portal/Mashup/Clients/Mast/Portal.html}} for Sector 19 in 2019 and for Sectors 43, 44, and 45 in the year 2021.  Each Sector has nearly continuous time series photometry with 120 second time resolution for a total of around 25 days (122 orbits).  In all, I have 89 eclipse times 1954--2021, plus four averaged eclipse times with each involving around 122 eclipses in each {\it TESS} Sector.

I have expended considerable effort in searching for old plates showing T Aur, in the hopes of extending the $O-C$ curve backwards in time.  I have searched and re-searched the archives at the Harvard, Sonneberg, Maria Mitchell, Asiago, Bamberg, and Vatican observatories.  This has produced 254 mags in 1929--1994 from Sonneberg, 2 mags in 1925 and 1936 from Maria Mitchell, 56 mags from Harvard in 1907--1955, plus 6 pre-eruption Harvard plates (1890--1891) for which the deepest limit is $B$$>$14.7.  Alas, these plates do not show any eclipses or orbital modulation, mainly because the exposure times (typically 45--60 minutes) are longer than the eclipse duration (with FWHM near 30 minutes) and the eclipse is only 0.2 mag deep.  Fortunately, these plates and magnitudes are good for measuring the century-long post-eruption light curve to seek any fading, as predicted by some models.

The uncertainties for the individual eclipse times have formal measurement errors (as in Table 2) that are always substantially smaller than the scatter in the $O-C$ curve.  The RMS scatters for the various individual observers varies with a median of 0.0019 days.  So the 0.0019 days is the intrinsic and irreducible jitter from the ordinary flickering of T Aur, variously making the times of minimum appear earlier or later than the times of conjunction.  The total uncertainties are from the addition in quadrature of the measurement errors and 0.0019 day jitter from flickering.  For each {\it TESS} Sector, with near 122 orbits of continuous photometry, the uncertainty contributed by the flickering will be $0.0019/\sqrt{122}$ = 0.00017 d, which when added to the measurement uncertainty (from the $\chi^2$ fit), results in a total uncertainty of near 0.00021 d.  These total uncertainties are used in the $\chi^2$ fits to a parabola, and used to calculate the weighted average for each observing season.

\begin{figure}
	\includegraphics[width=1.0\columnwidth]{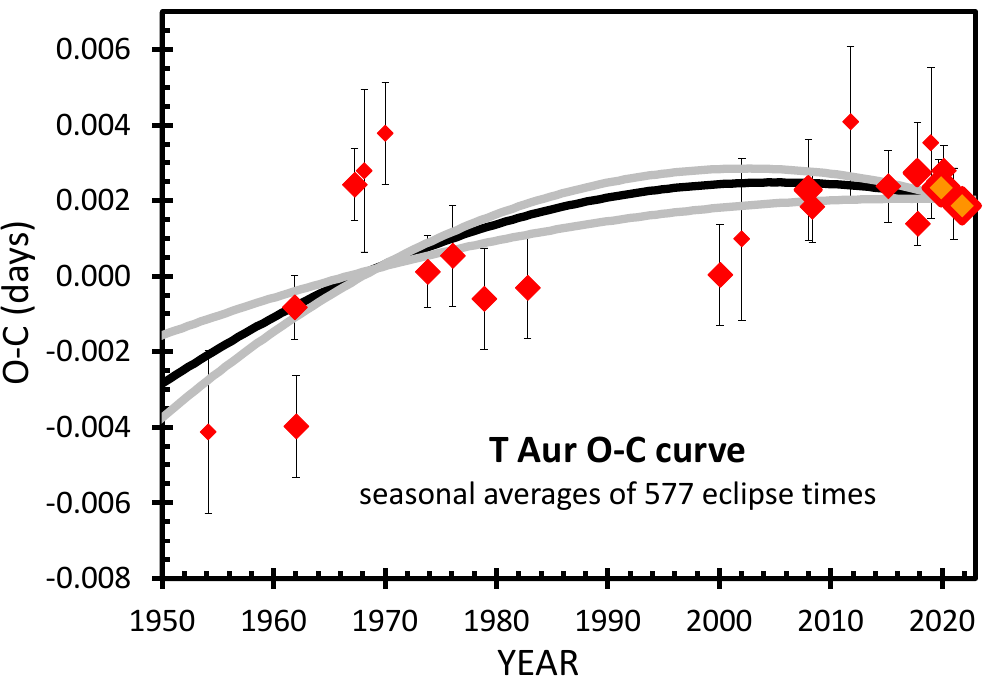}
    \caption{$O-C$ curve for T Aur. The seasonal averages are represented by the red diamonds, for a total of 89 individual eclipses plus 122 eclipses in each of four {\it TESS} Sectors.  The eclipse times from 2019 and 2021 produce $O-C$ measures that have very small error bars, with this issue being displayed as large red diamonds with an orange fill.  The eclipse times are well fit with a parabola, with a reduced $\chi^2$ near unity.  The parabola has a concave-down curvature (i.e., $\dot{P}$$<$0), which is to say that T Aur has a {\it decrease} in $P$.}
\end{figure}

All measured $O-C$  are collected in Table 2, and plotted in Fig. 2 as seasonal averages from the sources.  The `C' times were calculated from the linear ephemeris of Dai \& Qian (2010), with a period of 0.204378235 d and an epoch of HJD 2437614.0122.

The chi-square fits to the $O-C$ curve were calculated with all the individual eclipse times, except that the {\it TESS} points were fitted over all 122 orbits in each Sector.  The best-fitting parabola has an epoch of HJD 2437614.0114$\pm$0.0005, the period in 1961 is 0.20437832$\pm$0.00000005 days, and the dimensionless $\dot{P}$ is ($-$5.4$\pm$2.4)$\times$10$^{-12}$.  The $\chi^2$ for the no-curvature case is only 4.5 larger than the best case with curvature, so the existence of the $\dot{P}$ term as non-zero is only at the 2.1-sigma confidence level.  I note that the reduced $\chi^2$ for the best fitting parabola is sufficiently close to unity that any case is weak for systematic deviations from a parabola.  

\section{V394 CRA}

V394 CrA has known eruptions in the years 1949 and 1987, and has been little observed.  This RN peaked at $V$=7.2, faded with a plateau and $t_3$=5 days, hence a light curve class of P(5), reaching quiescence at $V$=18.4.  Schaefer (2009) discovered an orbital period of 1.515682 days, with a primary eclipse of 0.47 mag and a secondary eclipse of 0.26 mag.  I found that the out-of-eclipse brightness levels vary substantially, and the eclipse depths vary substantially.  With the observed $P$, the companion star must be an evolved sub-giant.

The orbital period can be determined from the eclipse times, but I have relatively few.  My light curves from Cerro Tololo observatory (CTIO) are distributed over nine observing seasons 1989--2008.  Further, {\it TESS} has excellent long time-series in Sectors 13 and 39 (2019 and 2021).  The folded light curve for Sector 39 is shown in Fig. 3.  (There is substantial uncertainty as to the background for subtraction in Fig. 3, due to not knowing the flux of the background stars in the relatively large photometry aperture, while the only result from this problem is that the amplitude in magnitudes is likely larger than would be derived from the displayed flux light curve.)  I have not been able to find any further useable light curves   These light curves make for 7 eclipse times (Table 3).

The $O-C$ curve is calculated with a linear ephemeris for a period of 1.515682 days and an epoch of HJD 2453660.81, as plotted in Fig. 3.  The $O-C$ curve can be fit to a straight line for no period change, or can be fit to a parabola for a steady period change.  My best-fitting parabola has a $\dot{P}$ value of ($-$5 $\pm$ 9)$\times$10$^{-10}$.  This value is consistent with a positive or negative or zero value.  This indeterminacy arises from my relatively few measures over only 32 years.

\begin{figure}
	\includegraphics[width=0.9\columnwidth]{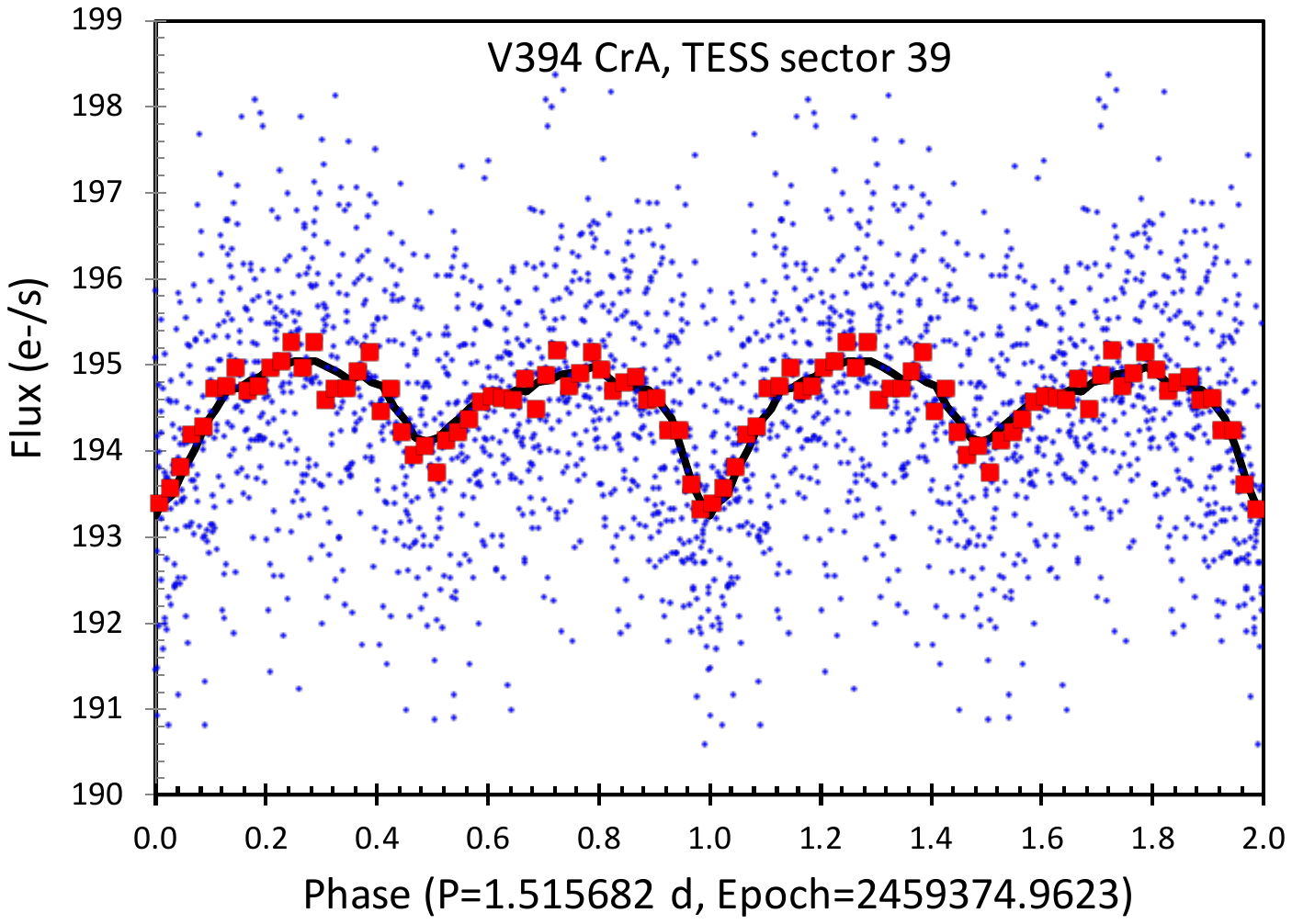}
    \caption{Folded light curve for V394 CrA.  The light curve from $TESS$ Sector 39 covers 21.8 days in middle-2021, with 600 second time resolution. This is folded about the orbital period to see the shape of the orbital variation, with the small blue dots indicating each flux measure.  The Poisson noise for each point is 1.3 e-/s, so much of the scatter in the folded light curve is ordinary noise.  The folded light curve has been binned into 50 phase bins, shown by the red squares.  These show a prominent primary minimum plus a secondary minimum.}
\end{figure}

\begin{table}
	\centering
	\caption{V394 CrA Eclipse Times}
	\begin{tabular}{lllrr}
		\hline
		Eclipse Time (HJD) & Source   &   Year   &   $N$   &   $O-C$ \\
		\hline
2447723.8730	$\pm$	0.0081	&	CTIO	&	1989.540	&	-3917	&	-0.0106	\\
2449904.9182	$\pm$	0.0101	&	CTIO	&	1995.511	&	-2478	&	-0.0318	\\
2453386.4769	$\pm$	0.0079	&	CTIO	&	2005.043	&	-181	&	0.0053	\\
2453910.9098	$\pm$	0.0093	&	CTIO	&	2006.479	&	165	&	0.0123	\\
2454574.7984	$\pm$	0.0094	&	CTIO	&	2008.297	&	603	&	0.0322	\\
2458670.1620	$\pm$	0.0052	&	TESS 13	&	2019.509	&	3305	&	0.0230	\\
2459374.9623	$\pm$	0.0044	&	TESS 39	&	2021.439	&	3770	&	0.0312	\\
		\hline
	\end{tabular}	
\end{table}

\begin{figure}
	\includegraphics[width=1.0\columnwidth]{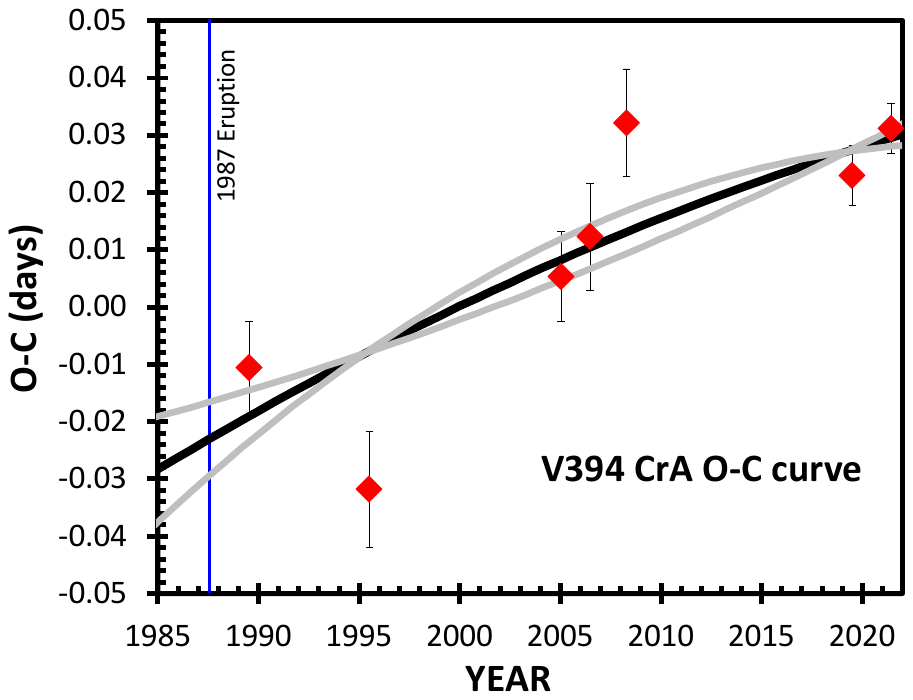}
    \caption{$O-C$ curve for V394 CrA.  I only have five eclipse times from Cerro Tololo and two recent eclipse times with {\it TESS} light curves.  The $O-C$ curve is poorly defined.  The best-fitting parabola (thick black curve) has only small curvature, while the one-sigma range for $\dot{P}$ (represented by the thick gray curves) go from positive to negative values.}
\end{figure}

\section{V1500 CYG}

V1500 Cyg (Nova Cyg 1975) was the brightest nova since 1946 (when T CrB peaked at $V$=2.0), and was discovered by many people high in the summer evening sky.  This nova has several unique and extreme properties, including a pre-eruption rise (from 21.5 to 13.5 $B$ mag in the month before eruption), a very large amplitude of outburst (19.5 mag), the then-fastest-known light curve speed ($t_3$=4 days), and that it's fading light curve is now 3.3 $B$ mag brighter than its pre-eruption level at a time 47 years after the eruption (hence becoming the prototype for the V1500 Cyg class of novae).  The orbital period was quickly determined from sinewave photometric modulations as 0.1396 days, while the spin period was measured from polarimetry to be slightly longer, hence V1500 Cyg became the prototype for a nova class called `asynchronous polars'.

I can construct an $O-C$ curve from 62 minimum times going back to 1976, see Table 4.  The fiducial ephemeris, for calculating the $O-C$, is from Semeniuk, Olech, \& Nale$\dot{\rm z}$yty (1995), with period of 0.13961296 days and an epoch of HJD 2446694.6730.  The first 50 times of minimum light have been taken from the literature from nine papers.  The minimum times from 1976 and 1977 display transient shifts unrelated to the orbital period changes (Patterson 1979, P1979) and are not used for the plots or fits of the $O-C$ curve.  (Similar transient shifts have been seen in the tails of the eruptions of U Sco and YZ Ret.)  The light curves from Sommers \& Naylor (1999, S\&N1999) and Harrison \& Campbell (2016, H\&C2016) were taken from a digitization of their figures.  These times of minimum are based on single orbits, where the omnipresent flickering shift the times of minimum substantially (Schaefer 2021).  These shifts can be quantified with the RMS of $O-C$ values for observations from a single observer and a single season, with values of near 0.0080 days.  

The databases of the AAVSO, ZTF, and {\it TESS} provide many magnitudes in light curves stretched out over many years.  I have extracted the light curves and then performed a chi-square fit of a sinewave to the light curve.  (These data show that the {\it average} light curve is close to a sinusoid.)  The time of minimum light for an epoch near the middle of the observing interval then represents the time of conjunction, with the one-sigma error bar coming from the range over which the $\chi^2$ value is within 1.0 of its minimum value.  I have used the light curves for an individual observing season, or similar group of data from an individual source, to produce minimum times.  The formal measurement error for the {\it TESS} data is small, from 0.0002--0.0005 days, but this does not include the uncertainty from the flickering jitter, which is 0.0080/$\sqrt{188}$ or 0.0006 days for the 188 orbits on average in a TESS Sector.


The V1500 Cyg $O-C$ curve in Fig. 5 shows a distinct and significant downward curvature.  The points before 1995 are flat by construction, because the fiducial ephemeris is from Semeniuk et al. (1995), while the 2007--2022 non-{\it TESS} points have significant negative slope.  This recent negative slope can also be separately determined from just the {\it TESS} minima times with their good real accuracy.  A formal $\chi^2$ fit to a parabola results in an epoch of HJD 2446694.6729$\pm$0.0004, a period in 1986 of 0.139612934$\pm$0.000000058 days, and a dimensionless $\dot{P}$ of ($-$2.7$\pm$1.0)$\times$10$^{-11}$.  This best-fitting parabola is shown as a thick black curve, while the one-sigma cases are shown as thick gray curves.  A fit to a straight line results in a $\chi^2$ close to 9 larger than the best-fitting parabola, so the existence of the curvature is near the 3-sigma confidence level.


\begin{table}
	\centering
	\caption{V1500 Cyg Minimum Times.  62 times in Supplementary Material}
	\begin{tabular}{lllrr}
		\hline
		Minimum Time (HJD) & Source   &   Year   &   $N$   &   $O-C$ \\
		\hline
2443640.9095	$\pm$	0.0080	&	P1979	&	1978.361	&	-21873	&	-0.0092	\\
2443665.9150	$\pm$	0.0080	&	P1979	&	1978.430	&	-21694	&	0.0056	\\
...   &  ...   &   ...   &   ...   &   ...   \\
2449994.5610	$\pm$	0.0080	&	S\&N1999	&	1995.756	&	23636	&	-0.0039	\\
2454393.6241	$\pm$	0.0080	&	H\&C2016	&	2007.800	&	55145	&	-0.0056	\\
2454394.6021	$\pm$	0.0080	&	H\&C2016	&	2007.803	&	55152	&	-0.0049	\\
2456875.7959	$\pm$	0.0080	&	H\&C2016	&	2014.596	&	72924	&	-0.0126	\\
2456900.7977	$\pm$	0.0080	&	H\&C2016	&	2014.665	&	73103	&	-0.0015	\\
2457634.0350	$\pm$	0.0046	&	AAVSO	&	2016.672	&	78355	&	-0.0115	\\
2457959.0451	$\pm$	0.0024	&	AAVSO	&	2017.562	&	80683	&	-0.0204	\\
2458340.6140	$\pm$	0.0021	&	ZTF	&	2018.607	&	83416	&	-0.0137	\\
2458466.6918	$\pm$	0.0057	&	ZTF	&	2018.952	&	84319	&	-0.0064	\\
2458731.6726	$\pm$	0.0044	&	ZTF	&	2019.677	&	86217	&	-0.0110	\\
2459058.6427	$\pm$	0.0036	&	ZTF	&	2020.573	&	88559	&	-0.0144	\\
2459131.0975	$\pm$	0.0025	&	AAVSO	&	2020.771	&	89078	&	-0.0188	\\
2459474.1285	$\pm$	0.0012	&	AAVSO	&	2021.710	&	91535	&	-0.0168	\\
2459736.0426	$\pm$	0.0017	&	AAVSO	&	2022.427	&	93411	&	-0.0166	\\
2458723.0107	$\pm$	0.0006	&	{\it TESS} 15	&	2019.654	&	86155	&	-0.0169	\\
2458750.0959	$\pm$	0.0007	&	{\it TESS} 16	&	2019.728	&	86349	&	-0.0166	\\
2459811.1510	$\pm$	0.0008	&	{\it TESS} 55	&	2022.633	&	93949	&	-0.0200	\\
		\hline
	\end{tabular}	
\end{table}

\begin{figure}
	\includegraphics[width=1.0\columnwidth]{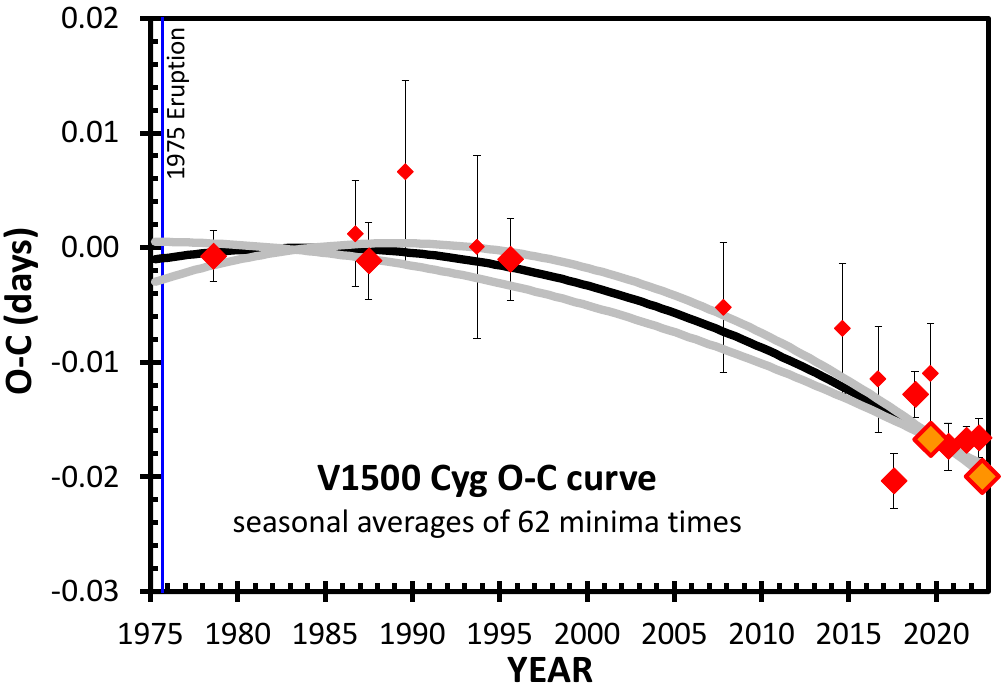}
    \caption{$O-C$ curve for V1500 Cyg.  This displays the deviations from the fiducial ephemeris (period of 0.13961296 days and epoch HJD 2446694.6730) for 62 times of minimum light, where seasonal averages are shown.  The {\it TESS} minima times from Sectors 15, 16, and 55 are shown as large orange diamonds, where the real error bars are greatly smaller than the the symbol.  The plot shows clear and significant curvature, with concave down, so the $P$ is decreasing. }
\end{figure}

\section{IM NOR}

\begin{table}
	\centering
	\caption{IM Nor Eclipse Times.  66 times in Supplementary Material.}
	\begin{tabular}{lllrr}
		\hline
		Eclipse Time (HJD) & Source   &   Year   &   $N$   &   $O-C$ \\
		\hline
2452696.5260	$\pm$	0.0005	&	WW2003	&	2003.154	&	0	&	-0.0005	\\
...   &  ...   &   ...   &   ...   &   ...   \\
2456152.5904	$\pm$	0.0007	&	CTIO 1.3m	&	2012.616	&	33674	&	0.0104	\\
2456166.0355	$\pm$	0.0005	&	SSO 2.3m	&	2012.653	&	33805	&	0.0106	\\
2458639.0973	$\pm$	0.0007	&	TESS 12	&	2019.424	&	57901	&	0.0348	\\
2458977.0733	$\pm$	0.0006	&	AAVSO	&	2020.349	&	61194	&	0.0413	\\
2458981.0749	$\pm$	0.0006	&	AAVSO	&	2020.360	&	61233	&	0.0402	\\
2458984.0512	$\pm$	0.0003	&	AAVSO	&	2020.368	&	61262	&	0.0402	\\
2459298.0095	$\pm$	0.0004	&	AAVSO	&	2021.228	&	64321	&	0.0451	\\
2459301.0891	$\pm$	0.0005	&	AAVSO	&	2021.236	&	64351	&	0.0457	\\
2459302.0122	$\pm$	0.0003	&	AAVSO	&	2021.239	&	64360	&	0.0452	\\
2459304.0654	$\pm$	0.0005	&	AAVSO	&	2021.245	&	64380	&	0.0457	\\
2459305.0912	$\pm$	0.0004	&	AAVSO	&	2021.247	&	64390	&	0.0451	\\
2459307.0408	$\pm$	0.0003	&	AAVSO	&	2021.253	&	64409	&	0.0447	\\
2459313.0980	$\pm$	0.0003	&	AAVSO	&	2021.269	&	64468	&	0.0466	\\
2459314.0201	$\pm$	0.0004	&	AAVSO	&	2021.272	&	64477	&	0.0450	\\
2459338.0334	$\pm$	0.0009	&	TESS 38	&	2021.338	&	64711	&	0.0423	\\
2459368.0051	$\pm$	0.0010	&	TESS 39	&	2021.420	&	65003	&	0.0452	\\
		\hline
	\end{tabular}	
\end{table}

\begin{figure}
	\includegraphics[width=1.0\columnwidth]{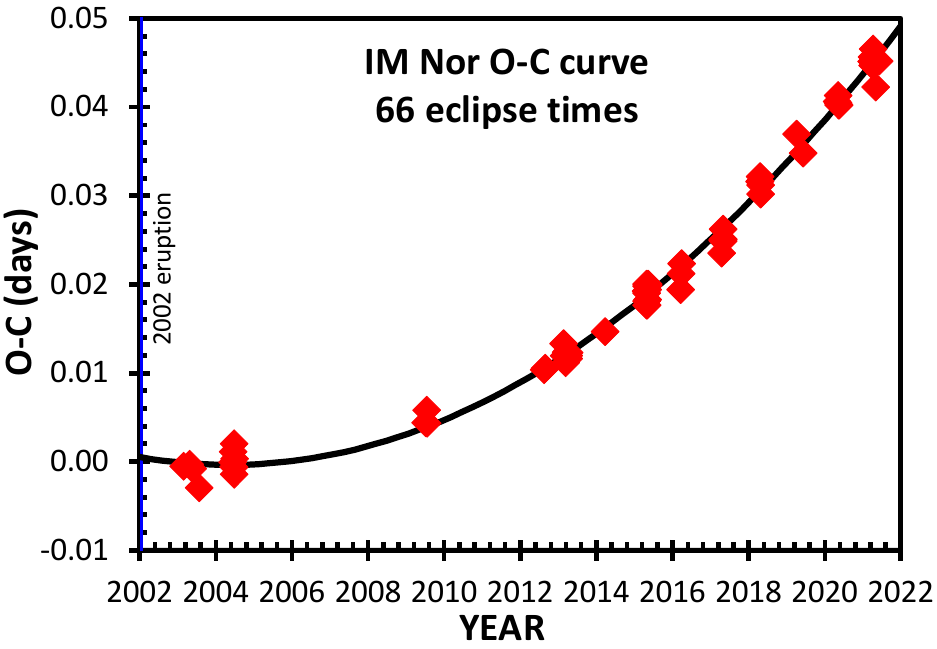}
    \caption{$O-C$ curve for IM Nor.  The 66 eclipse times from 2003--2021 (see Table 5) form an $O-C$ curve as based on the fiducial ephemeris with an epoch of HJD 2452696.5265 and a period of 0.1026327 days.  These times form an $O-C$ curve that is a good parabola with $\dot{P}$ equal to ($+$2.51$\pm$0.07)$\times$10$^{-11}$.  The best-fitting parabola (thick black curve) is concave-up, meaning that the period is {\it increasing}.}
\end{figure}

IM Nor has RN eruptions observed in 1920 and 2002, although several intermediate eruptions were likely missed.  The eruptions peaked at $V$=8.5, faded by three magnitudes in 80 days, displayed a plateau in the light curve for class P(80), returning to a quiescent level of $V$=18.3.  Woudt \& Warner (2003; WW2003) discovered the orbital period of 0.1026 days, with the folded light curve showing a broad primary eclipse with a broad secondary eclipse.  Startlingly, IM Nor is inside the `Period Gap'.  The large $t_3$ value (large for a RN), the light curve shape, and the very short P are similar to T Pyx.  

With IM Nor inside the Period Gap, the MBM requires that the accretion rate be very low, under 10$^{-10}$ M$_{\odot}$ yr$^{-1}$ (Knigge et al. 2011).  But to accumulate enough mass within 82 years or less to trigger a nova eruption (i.e., to be a RN), the accretion rate must be near 10$^{-7}$ M$_{\odot}$ yr$^{-1}$ (Shen \& Bildsten 2009).  This sets up a severe dilemma for standard CV evolution models, where such a high accretion rate from such a short-$P$ system is impossible.  To save the case, there must be some unknown and unmodelled mechanism that is currently driving $\dot{M}$ to be $>$1000$\times$ larger than expected.  Following the case of its sister-RN T Pyx, I speculate that IM Nor now has a long-lasting yet transient episode of nuclear burning on the surface of the white dwarf (triggered by a CN event over a century ago) that is irradiating the companion star's surface so as to drive a very high $\dot{M}$ and RN eruptions coming every few decades.

As a measure of this evolution, Patterson et al. (2022) has a collection of 49 eclipse times from 2003 to 2020 and derive a dimensionless $\dot{P}$ of $+$2.18$\times$10$^{-11}$.  I have extended the $O-C$ curve of Patterson et al. (2022) by adding many eclipse times.  Two of these are times I observed in 2012, at Cerro Tololo observatory (CTIO with the 1.3-m telescope) and at Siding Spring Observatory (SSO with the 2.3-m telescope).  Further, I have used the {\it TESS} light curves for Sectors 12, 38, and 39, where I have fitted a template to the observed light curve, resulting in an eclipse time averaged over $\sim$250 orbits each Sector and hence of higher accuracy than from single-orbit photometry.  Further, I have collected the light curves of G. Myers, as reported in the AAVSO database, and fitted parabolas to the minima around the times of eclipse, for  11 eclipse times 2020--2021.  In all, I have 16 new eclipse times, including three times of high accuracy from {\it TESS}, based on roughly 760 eclipses.  All the eclipse times (as heliocentric Julian dates) are quoted in Table 5.  The eclipse times reported in Patterson et al. (2022) appear only in the Supplementary Material version of Table 5.


My 66 eclipse times were fitted to a quadratic model, with the best-fitting parabola displayed in Fig. 6.  Again, the quadratic term is $+0.5 P \dot{P} N^2$, so $\dot{P}$ has units of days per day, or dimensionless.  I find $\dot{P}$ equal to  ($+$2.51$\pm$0.07)$\times$10$^{-11}$.  (The period and epoch for $N$=0 are 0.102632593 $\pm$ 0.000000025 days and HJD 2452696.52636 $\pm$ 0.00024.)  For the parabola fit, I get a reduced $\chi^2$ near unity only if I add in quadrature a systematic error of 0.0010 days, which is to say that real measurement errors plus the intrinsic jitter in eclipse minima times is 1.4 minutes, with this not included in the formal uncertainties quoted in Table 5.

\section{T PYX}

T Pyx is a recurrent nova with observed eruptions in 1890, 1902, 1920, 1944, 1967, and 2011, reaching a peak magnitude of $V$=6.4 and then fading three magnitudes in 62 days.  I first recognized its orbital period as 0.07616 d (Schaefer et al. 1992), but it was Patterson et al. (1998) who first worked out the confident and accurate period (0.0762233 d), the light curve, and even the fast period changes for $\dot{P}$.  T Pyx has a light curve with a very broad primary eclipse, plus a shallow secondary eclipse, plus flickering and long-term variations.  With a period of 1.83 hours, T Pyx is in the case where it is inside the nova Period Gap\footnote{For CVs, the Period Gap is the range of $P$ over which few binaries occur, traditionally taken to be 2--3 hours.  However, novae have a distinct `gap' from 1.70--2.66 hours (Schaefer 2022b), so T Pyx is actually inside the Period Gap.}.  This sets up the paradox as to how it is possible that a binary inside the Period Gap can possibly have such a large accretion rate as to allow frequent RN events.

T Pyx has a nova shell consisting of over 30 unresolved  bright knots.  Schaefer, Pagnotta, \& Shara (2010) used {\it Hubble Space Telescope} images from 1994--2007 to measure the expansion rate of the knots as 500--715 km s$^{-1}$, and derived that the year of the eruption that emitted the knots was close to 1866.  The total mass of the 1866 ejecta is $\sim$10$^{-4.5}$ M$_{\odot}$, which is larger than possible from any of the T Pyx RN eruptions.  The expansion velocity and shell mass prove that the 1866 eruption was not an RN event.  Rather, the 1866 eruption was a classical nova eruption, where the large ejecta mass forces a long accumulation time interval before 1866, presumably at the expected low accretion rate.  Before 1866, T Pyx was some sort of ordinary CV with a short period (but with a near Chandrasekhar mass white dwarf) that finally accumulated enough mass to trigger an ordinary nova eruption that ejected the knots.  Subsequent RN event ejected a high-velocity low-density wind that sculpted the 1866 ejecta into knots by the Rayleigh-Taylor instability.    The 1866 classical nova eruption kickstarted the system into a century-long state of very-high accretion.  This unique high-state is demonstrated by the luminous quiescent level, where for example the brightness {\it before} the 1890 RN event was $B$=13.8, which corresponds to an absolute $B$ magnitude of $+$0.2, with this being brighter than almost all known CVs.  This high accretion rate plus the high white dwarf mass are what allowed T Pyx to have fast recurrence time-scales for thermonuclear runaway eruptions that eject relatively small masses with high velocity.  The high-accretion state started by the 1866 eruption has been fading away.  This secular decline in the accretion rate is seen by the quiescent $B$ magnitude fading from 13.8 in 1890, to 14.3 in 1900, to 14.9 in 1930, to 15.5 in 1980, and then to 16.1 in 2022.  The secular decline in accretion rate is also seen in the inter-eruption time intervals, lengthening from 11.9, to 17.9, to 24.6, to 22.1, and then to 44.3 years.  After more than a century in a high-state, T Pyx is slowly returning to its pre-1866 low-accretion state.  Currently, the accretion rate is sufficiently small that I expect no more RN events for many millennia.  The primary science question is trying to understand the nature of the transient high-accretion state that was kickstarted by the ordinary classical nova eruption of 1866.  Presumably, T Pyx separates itself from most other CVs by its very short $P$ and by having a white dwarf near the Chandrasekhar mass.  The general case where novae keep a state of high accretion, lasting for many decades after the eruption has stopped, is known for seven other systems, now labelled as `V1500 Cyg stars' (Schaefer \& Collazzi 2010).  T Pyx is the most extreme known case of the V1500 Cyg transient-high-state phenomenon.  The physical mechanism is likely somehow associated with irradiation from the hot white dwarf (after the nova eruption) puffing up the atmosphere of the companion star so as to pump up the accretion rate.

Patterson et al. (1998) and Schaefer et al. (2013, S2013) measured $\dot{P}$ before the 2011 eruption, while Patterson et al. (2017) measured the $\Delta P$ across the 2011 eruption.  The $O-C$ curve of Patterson and colleagues includes 77 eclipse times from 1996--2016.  Here, I can extend the $O-C$ curve both backwards and forwards in time, so as to cover 1986--2022.  I add my 19 eclipse times from 1986-2011, all observed with telescopes at Cerro Tololo observatory, as listed in Schaefer et al. (2013).  Further, I have extracted time series from the AAVSO International Database for 2011--2022, and fitted the folded light curves from collections of nights to a T Pyx template so as to derive the times of minima.  The individual observers are S. Dvorak, F.-J. Hambsch, L. Monard, G. Myers, P. Nelson, and A. Oksanen, all amongst the most-experienced CV observers in the world, and associated with the Center for Backyard Astronomy (CBA; J. Patterson P. I.).  In all, I have 31 eclipse times from CBA observers.  Further, I have taken the {\it TESS} light curves for Sectors 8 (in 2019) and 35 (in 2021) and fitted a light curve template so as to derive an averaged time of minimum.  Each {\it TESS} Sector is near 24 days in duration, covering over 310 orbits for averaging.  In all, I have 129 eclipse times 1986--2022.  These are presented in Table 6, although the previously published values in Patterson et al. (1998) and Schaefer et al. (2013) appear only in the Supplementary Material.

\begin{table}
	\centering
	\caption{129 T Pyx Eclipse Times (full table in Supplementary Material)}
	\begin{tabular}{lllrr}
		\hline
		Eclipse Time (HJD) & Source   &   Year   &   $N$   &   $O-C$ \\
		\hline
2446439.5121	$\pm$	0.0035	&	S2013	&	1986.023	&	-121040	&	0.2934	\\
2447141.0770	$\pm$	0.0056	&	S2013	&	1987.944	&	-111836	&	0.2451	\\
...			&	...	&	...	&	...	&	...	\\
2455830.0506	$\pm$	0.0010	&	AAVSO	&	2011.733	&	2152	&	0.0092	\\
2455865.7979	$\pm$	0.0010	&	AAVSO	&	2011.831	&	2621	&	0.0051	\\
2455866.0315	$\pm$	0.0009	&	AAVSO	&	2011.832	&	2624	&	0.0100	\\
2455923.1284	$\pm$	0.0007	&	AAVSO	&	2011.988	&	3373	&	0.0112	\\
2456030.0096	$\pm$	0.0006	&	AAVSO	&	2012.281	&	4775	&	0.0192	\\
2456200.0131	$\pm$	0.0004	&	AAVSO	&	2012.746	&	7005	&	0.0316	\\
2457426.7751	$\pm$	0.0026	&	AAVSO	&	2016.105	&	23097	&	0.1140	\\
2457428.8263	$\pm$	0.0016	&	AAVSO	&	2016.110	&	23124	&	0.1070	\\
2457431.0368	$\pm$	0.0006	&	AAVSO	&	2016.116	&	23153	&	0.1069	\\
2457735.7514	$\pm$	0.0036	&	AAVSO	&	2016.951	&	27150	&	0.1335	\\
2457739.0293	$\pm$	0.0008	&	AAVSO	&	2016.960	&	27193	&	0.1336	\\
2457835.0072	$\pm$	0.0013	&	AAVSO	&	2017.222	&	28452	&	0.1389	\\
2458101.8273	$\pm$	0.0013	&	AAVSO	&	2017.953	&	31952	&	0.1570	\\
2458153.1319	$\pm$	0.0005	&	AAVSO	&	2018.093	&	32625	&	0.1594	\\
2458463.1016	$\pm$	0.0008	&	AAVSO	&	2018.942	&	36691	&	0.1813	\\
2458529.0452	$\pm$	0.0023	&	TESS 8	&	2019.123	&	37556	&	0.1867	\\
2458546.0453	$\pm$	0.0003	&	AAVSO	&	2019.169	&	37779	&	0.1877	\\
2458585.8411	$\pm$	0.0011	&	AAVSO	&	2019.278	&	38301	&	0.1918	\\
2458888.7247	$\pm$	0.0016	&	AAVSO	&	2020.107	&	42274	&	0.2170	\\
2458909.0768	$\pm$	0.0003	&	AAVSO	&	2020.163	&	42541	&	0.2159	\\
2458919.0564	$\pm$	0.0009	&	AAVSO	&	2020.190	&	42672	&	0.2095	\\
2459202.1245	$\pm$	0.0008	&	AAVSO	&	2020.965	&	46385	&	0.2387	\\
2459238.8682	$\pm$	0.0004	&	AAVSO	&	2021.066	&	46867	&	0.2400	\\
2459240.1633	$\pm$	0.0005	&	AAVSO	&	2021.070	&	46884	&	0.2392	\\
2459249.6965	$\pm$	0.0016	&	AAVSO	&	2021.096	&	47009	&	0.2437	\\
2459268.0687	$\pm$	0.0009	&	TESS 35	&	2021.146	&	47250	&	0.2447	\\
2459593.5916	$\pm$	0.0015	&	AAVSO	&	2022.037	&	51521	&	0.1928	\\
2459619.7330	$\pm$	0.0011	&	AAVSO	&	2022.109	&	51863	&	0.2639	\\
2459639.7095	$\pm$	0.0005	&	AAVSO	&	2022.163	&	52126	&	0.1921	\\
2459644.3609	$\pm$	0.0012	&	AAVSO	&	2022.176	&	52187	&	0.1935	\\
2459658.6941	$\pm$	0.0015	&	AAVSO	&	2022.215	&	52375	&	0.1956	\\
2459666.3209	$\pm$	0.0015	&	AAVSO	&	2022.236	&	52475	&	0.1995	\\
2459672.7202	$\pm$	0.0011	&	AAVSO	&	2022.254	&	52559	&	0.1956	\\
		\hline
	\end{tabular}	
\end{table}

\begin{figure}
	\includegraphics[width=1.0\columnwidth]{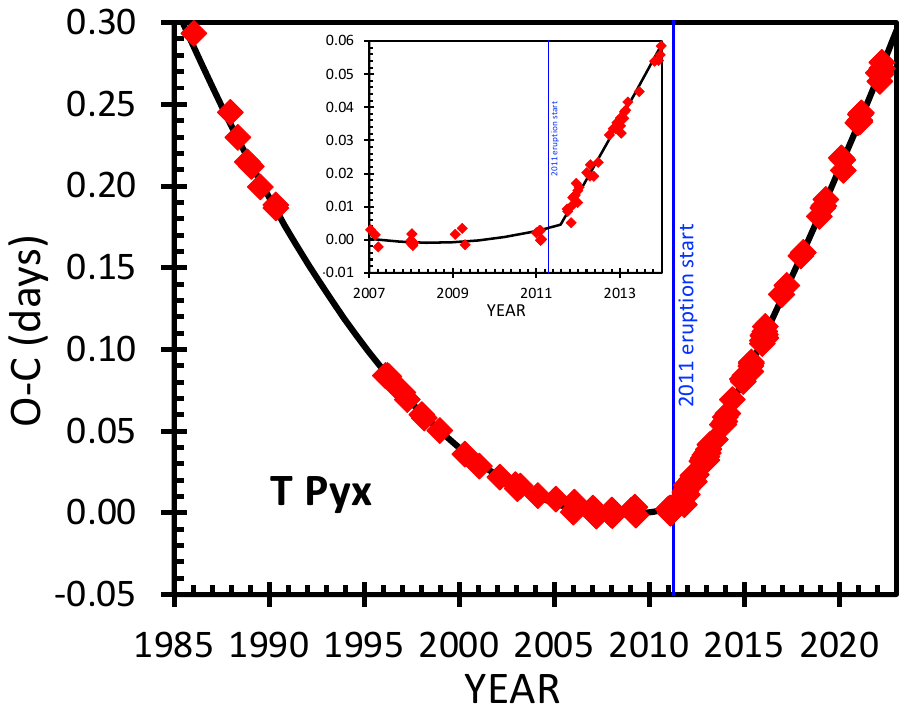}
    \caption{$O-C$ curve for T Pyx.  The 129 minima times (red diamonds) are closely fit to a broken parabola (the thick black curve).  The sharply defined break occurs 108.4$\pm$12.6 days after the start of the nova event (indicated by the thin blue vertical line), as seen in the inset showing the years around the 2011 eruption.  In quiescence, the parabolas are concave-up, meaning that the orbital period is increasing, with the rate of increase consistent with that expected for the high mass transfer in a RN.  The sharp kink in the $O-C$ curve is caused by a sudden period increase, by $+$50.3 ppm. }
\end{figure}

For the $O-C$ values, I choose a fiducial linear ephemeris with an epoch of HJD 2455665.9962 (near the start of the 2011 eruption) and a period of 0.07622918 d.  The calculated year, $N$, and $O-C$ for each minimum time is given in Table 6.  The $O-C$ curve is plotted in Fig. 7.  We see a good parabola, concave up, before the 2011 eruption.  We see a sharp upward kink around the time of the eruption.  We see a good parabola, concave up, after the nova event.

The pre-eruption $O-C$ is fit to a parabola (Eq. 1) with $E_0$ at HJD 2455665.99973$\pm$0.00043, $P_{\rm pre}$=0.076229825$\pm$0.000000025 d, and $\dot{P}_{\rm pre}$ equals ($+$6.49$\pm$0.07)$\times$10$^{-10}$.  The post-eruption $O-C$ is fit to a parabola with zero-epoch of HJD 2455665.9943	$\pm$0.0004, $P_{\rm post}$=0.07623366$\pm$0.00000060 d, and $\dot{P}_{\rm post}$ of ($+$3.67$\pm$0.27)$\times$10$^{-10}$.  The period change across the eruption is $\Delta P$=0.00000383$\pm$0.00000060 d, for which the fractional period change is $+$50.3$\pm$7.9 ppm.  The RMS scatter of the $O-C$ values around this best-fitting broken parabola is 0.0014 d above the formal measurement errors, and this is the random jitter in times due to the ordinary flickering.  The fits to the parabolas are good, meaning that the $\dot{P}$ is close to constant from 1986--2011 and then 2011--2022.

The best fit to a broken parabola does {\it not} have the break at $N$=0 (i.e., the time of the start of the eruption).  This is readily seen in the inset to Fig. 7, where the post-eruption parabola crosses $N$=0 for an $O-C$ significantly before the time when the post-eruption parabola crosses $N$=0.  The intersection of the pre- and post-eruption parabolas (i.e., the time of the break) is for $N$=1422, a total of 108.4$\pm$12.6 days after the start of the eruption.  The best fit that force the break to occur at $N$=0 has a $\chi^2$ value 74.2 larger than the unconstrained case, which is to say that the break is significantly later than the start of the eruption.  

The $\Delta P$ mechanisms act nearly instantaneously with the ejection of mass from the binary.  So my observations say that the average time of ejection is the 108.4 days after the start of the eruption.  So the time profile for mass ejection must have at least half the ejection after 108.4 days.  T Pyx now provides the first measure of the time distribution of the shell formation, and hopefully future theory models can provide some physical understanding.

\begin{figure}
	\includegraphics[width=1.0\columnwidth]{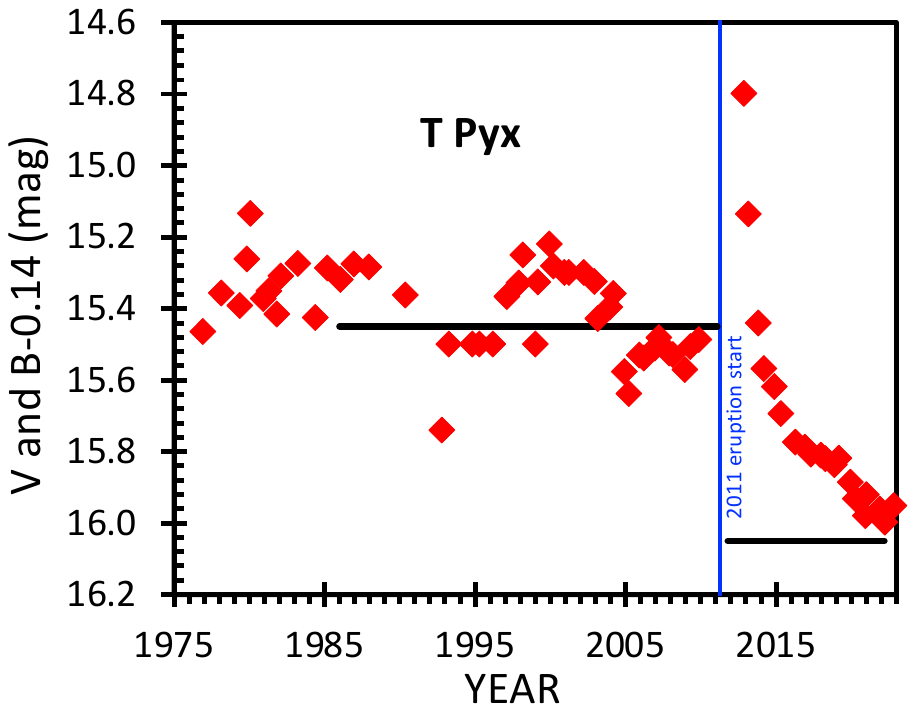}
    \caption{T Pyx in quiescence 1976--2022.  The AAVSO light curve is binned variously to half-year and one-year time bins.  All the magnitudes are with CCD and with the modern comparison sequence.  The magnitudes before 1990 are $B$ mags as corrected to $V$ with the stable $\langle B-V \rangle$=0.14 mag (Schaefer et al. 2013; Bruch \& Engel 1994).  The start date of the 2011 nova event is indicated with the thin blue vertical line.  The $O-C$ curve has constant $\dot{P}$ from 1986--2011 and 2011--2022, which forces $\dot{M}$ and $V$ to be constant from 1986--2011 and a different constant from 2011--2022.  Independently derived from $\Delta P$ and from $\dot{P}_{pre}$/ $\dot{P}_{post}$, the quiescent magnitude should drop by roughly 0.6 mag from 1986--2011 to 2011--2022.  This is illustrated by the two horizontal, thick, black lines, with $V$=15.45 before the eruption and $V$=16.05 after the eruption.  The pre-eruption interval matches a constant level to within the ordinary variations seen in all novae.  The post-eruption interval shows the fading from the nova event, with the light curve asymptoting to $V$=16.05.}
\end{figure}

\section{Updates on Period Changes for Eight Novae}

{\bf QZ Aur} was a poorly observed nova in 1964, and declined to a quiescent level near $V$=17.0.  Campbell \& Shafter (1995) discovered 10 eclipses from 1990--1992 with depth of 1.3 mag for a period of  0.357496 days.  Shi \& Qian (2014) added 8 eclipse times 2008--2013.  Schaefer et al. (2019) extended the eclipse times back to 1952 with archival plates from the Vatican, Sonneberg, and Asiago observatories, and measuring the sudden orbital period change across the 1964 eruption.  Further, 13 eclipse times were added from 2009--2016, allowing a good measure of the post-eruption $\dot{P}$.  Here, I update the $O-C$ curve by adding mean eclipse times for four $TESS$ Sectors from 2019 and 2021.  The new eclipse times are HJD 2458828.52116$\pm$0.00018, 2459486.31459$\pm$0.00014, 2459512.05427$\pm$0.00018, and 2459538.15142$\pm$0.00016.  With these, the updated value for the post eruption $\dot{P}$ is ($-$3.9 $\pm$ 1.4)$\times$10$^{-11}$.  The possibility of a zero-$\dot{P}$ is rejected at the 3.0-sigma confidence level.  The new $\dot{P}$ is negative, implying a steadily {\it decreasing} orbital period.    

{\bf T CrB} is a recurrent nova with eruptions in 1866 and 1946, and it is widely expected that it will have another eruption in 2025.5$\pm$1.3 (Schaefer 2023).  The eruptions reach a peak at $V$=2.0, with this being the brightest nova event since 1943.  The companion star is a red giant, M4 {\rm III}, so the orbital period is 227.5687$\pm$0.0099 days, and a prominent ellipsoidal modulation appears with typical full-amplitude of 0.3 mag at half the orbital period (Kenyon \& Garcia 1986; Leibowitz, Ofek, \& Mattei 1997; Fekel et al. 2000).  I have collected 213,730 $B$ and $V$ magnitudes from 1842--2022, and this light curve closely defines the phase of the ellipsoidal variations all the way back to 1867 (Schaefer 2023).  The ellipsoidal minima can be used to construct an $O-C$ curve that tracks the changes in the orbital period from 1867--2022.  The resultant $O-C$ curve displays a significant broken parabola, with the break in 1946.  From this, the $\dot{P}$ is measured for 1867--1946 and 1947--2022, plus the sudden change in $P$ across the 1946 eruption is measured.  The $\dot{P}$ changed from near-zero ($+$1.75$\pm$4.5)$\times$10$^{-6}$) before 1946 to a large negative value ($-$8.9$\pm$1.6)$\times$10$^{-6}$ after 1946.   The orbital period {\it increased} by 0.185$\pm$0.056 days in 1946.  These $P$ changes cannot be explained by any published model.

{\bf HR Del} was a bright nova, peaking in 1967 at $V$=3.6 mag, with the light curve filled with jitters up and down (J-class), while the decline was rather slow with $t_3$=231 days (Schaefer 2022c).  In quiescence, HR Del is one of the brightest novae at $V$=12.1.  The nova has a sinusoidal modulation with full-amplitude near 0.10 mag, for an orbital period of 0.214 days,  Schaefer (2020b) used archival plates from Harvard and Sonneberg observatories to measure the pre-eruption $P$ and $\dot{P}$, plus the $\Delta P$ for the 1967 eruption.  The post-eruption period and $\dot{P}$ was found from 7 timed phases from 2004--2017.  As a small update, I can now add the time of maximum light to be HJD 2459811.0457$\pm$0.0002 for {\it TESS} Sector 55 in middle 2022.  The 2004--2022 times are not long enough to see any curvature in the {\it O-C} curve, but curvature is forced so as to reach the epoch for the start of the post-eruption interval at the time of the eruption as determined from the pre-eruption {\it O-C} curve.  The slightly updated $\dot{P}$ is ($+$1.63$\pm$0.37)$\times$10$^{-10}$.

{\bf DQ Her} is one of the most famous, nearest, and brightest novae of all time.  It peaked at $V$=1.6 in late 1934, faded with $t_3$=100 days, displayed the prototypical dust dip D-class light curve, and presented a bright expanding shell that is still easily visible.  In quiescence, DQ Her is near $V$=14.4, and is now known as the prototype of the Intermediate Polars where their highly magnetized white dwarf funnels material in the inner accretion disc.  In 1954, M. F. Walker discovered that DQ Her was in a binary system that eclipses with a period near 0.1936 days (Walker 1954), and this started the real understanding of the nature of nova eruptions (Kraft 1958; 1959).  With a moderately bright system and well-defined deep eclipses, many workers have reported 144 eclipse times from 1954 to 2018 (Schaefer 2020a).  The resultant $O-C$ curve is basically flat with nominally-significant small-and-fast fluctuations, for which prior workers have tried fitting one and two sinewaves.  Schaefer (2021) demonstrated that these variations are just timing noise resulting from ordinary flickering shifting the apparent times of minima.  Importantly, Schaefer (2020a) measured 52 pre-eruption $B$ magnitudes (1894--1934) from the Harvard and Sonneberg plates, with three eclipse times, resulting in an accurate pre-eruption $P$ and $\Delta P$.  For the post-eruption $O-C$ curve, Schaefer (2020a) used 144 eclipse times (1954--2018) to return a $\dot{P}$ close to zero.  Now, I can update this post-eruption analysis with eclipse times from 5 {\it TESS} Sectors (2020--2022) with 120 second time resolution.  Each Sector has an average of 119 eclipses well-measured, so these eclipse times have a high formal accuracy and the jitter from flickering has been reduced by an order-of-magnitude.  For each Sector, I have fitted a light curve template to the phase curves, so as to produce a single average eclipse time near the middle of the Sector.  These five eclipse times are HJD 2458997.045854$\pm$0.000027, 2459023.184690$\pm$0.000026, 2459405.199208$\pm$0.000018, 2459732.030489$\pm$0.000021, and 2459783.146470$\pm$0.000026.  With these new exacting eclipse times, I have fit the 149 post-eruption eclipse times to a simple parabola in the $O-C$ curve.  The updated $\dot{P}$ value is ($+$1.3$\pm$1.2)$\times$10$^{-13}$.  This steady period change is very small, and is consistent with zero.

{\bf BT Mon} was discovered in 1939 with the Harvard plates, and it displayed a flat-top peak (F-class) lasting for over 75 days near $V$=8.1 (Schaefer 2020a).  Robinson, Nather \& Kepler (1982) discovered deep eclipses with a period close to one-third of a day.  These eclipses are deep and relatively long, so they are readily seen on archival plates, even with the quiescent nova at $B$=15.7 usually being near the plate limit.  With this, Schaefer \& Patterson (1983) found many pre-eruption plates showing BT Mon, with a number of eclipses, hence measuring the $P_{pre}$ and $\Delta P$.  With the orbital period {\it increasing} across the eruption, and the orbital separation necessarily getting larger suddenly, Shara et al. (1986) took BT Mon as part of their inspiration for the Hibernation Model of CV evolution.  Schaefer (2020a) presented a substantially improved data set, and reported a similar, but much more accurate, $\Delta P$.  Now, I can add a further update with a Sector of {\it TESS} data ending in January 2021, containing 73 eclipses well-measured with 20-second time resolution.  The phase-averaged eclipse time is HJD 2459203.64498$\pm$0.00003.  I have derived a slightly-improved post-eruption $\dot{P}$ to be ($-$6.80$\pm$0.30)$\times$10$^{-11}$.

{\bf RR Pic} peaked in middle-1925 at $V$=1.0, making it the third-brightest known novae.  With a distance of 501 pc, the nova is one of the closest known novae.  In quiescence, RR Pic is at $V$=12.2 and displays quasi-periodic oscillations and superhumps (Schmidtobreick et al. 2008).  Van Houten (1966) discovered a roughly sinewave modulation with a period near 0.145 days.  Vogt et al. (2017) has collected 203 times of maximum from 1965--2014.  I extended the $O-C$ curve back to 1945 and forward to 2017, finally being able to see a significant and good parabola so as to measure the $\dot{P}$ (Schaefer 2020b).  Further, I measured 82 $B$ magnitudes with the Harvard plates from 1889 to 1925, so as to derive the pre-eruption $P$ as well as $\Delta P$.  Now, I have derived 22 times of maximum light from 22 Sectors of {\it TESS} data (2018--2021), always with a formal chi-square fit to a sinewave, as listed in Table    7.  The {\it TESS} light curves show a stable shape, with a nearly flat maximum over 0.50 in phase, plus a nearly-flat minimum lasting 0.20 in phase that consistently shows shallow dips separated by 0.14 in phase.  In this situation, it becomes difficult to define a phase of maximum light.  For purposes of the $O-C$ diagram, the important issue is that all observers use the same effective definition.  Fortunately, all measures are either fits to a sinewave or with a method invoking symmetry that should produce similar times.  Nevertheless, uncertainties involving the derivations of maxima times in the light curves might add small amounts of scatter into the $O-C$ curve.  The 22 {\it TESS} times have an average formal measurement error bar of 0.00015 days (13 seconds), and an RMS of their $O-C$ values of 0.0037 days (320 seconds).      The RMS of the $O-C$ deviations from the best-fitting parabola is 0.0024 days (210 seconds) for the Harvard archival data, and the RMS of the deviations is 0.0078 days (675 seconds) for all the non-Harvard and non-{\it TESS} times.  The new {\it TESS} $O-C$ measures closely straddle the best-fitting parabola from Schaefer (2020b), with an average difference of 0.00057 (49 seconds).  With my updated $O-C$ curve, I have fitted a parabola to all the post-eruption times, and I derive an updated $\dot{P}$ value of ($+$9.58$\pm$0.34)$\times$10$^{-11}$.

\begin{table}
	\centering
	\caption{RR Pic times of maximum light with {\it TESS}}
	\begin{tabular}{ll}
		\hline
		{\it TESS} Sector & Time (HJD)   \\
		\hline
{\it TESS} 2	&	2458367.05627	$\pm$	0.00014	\\
{\it TESS} 3	&	2458396.06223	$\pm$	0.00016	\\
{\it TESS} 4	&	2458424.05301	$\pm$	0.00016	\\
{\it TESS} 5	&	2458451.02776	$\pm$	0.00014	\\
{\it TESS} 6	&	2458479.01727	$\pm$	0.00015	\\
{\it TESS} 7	&	2458504.10650	$\pm$	0.00014	\\
{\it TESS} 8	&	2458529.05057	$\pm$	0.00016	\\
{\it TESS} 9	&	2458556.02622	$\pm$	0.00015	\\
{\it TESS} 10	&	2458584.01599	$\pm$	0.00016	\\
{\it TESS} 12	&	2458641.15582	$\pm$	0.00015	\\
{\it TESS} 13	&	2458670.01692	$\pm$	0.00014	\\
{\it TESS} 27	&	2459047.08403	$\pm$	0.00015	\\
{\it TESS} 29	&	2459100.01868	$\pm$	0.00015	\\
{\it TESS} 30	&	2459128.00900	$\pm$	0.00014	\\
{\it TESS} 31	&	2459157.01370	$\pm$	0.00014	\\
{\it TESS} 32	&	2459180.07383	$\pm$	0.00019	\\
{\it TESS} 33	&	2459215.02468	$\pm$	0.00013	\\
{\it TESS} 34	&	2459242.14452	$\pm$	0.00013	\\
{\it TESS} 35	&	2459268.10438	$\pm$	0.00015	\\
{\it TESS} 36	&	2459294.06399	$\pm$	0.00014	\\
{\it TESS} 37	&	2459321.03871	$\pm$	0.00014	\\
{\it TESS} 39	&	2459376.14741	$\pm$	0.00013	\\

		\hline
	\end{tabular}	
\end{table}

{\bf V1017 Sgr} was a poorly observed nova in 1919, only caught late in the tail.  The nova faded to $V$=13.5, although dwarf nova eruptions have been recorded in 1901, 1973, and 1991, with half-year triangular-shaped light curves (Salazar et al. 2017).  Sekiguchi (1992) presented nine radial velocities, found a G-star spectrum, and selected from the many equal aliases a periodicity of $\sim$5.7 days.  V1017 Sgr is one of the brightest of all novae in quiescence, so it is surprising that no follow-up studies were made, and knowledge of V1017 Sgr was still sparse.  Finally, in 2017, Salazar et al. (2017) made a full long-term study of all photometry, including the light curve from the Harvard plates from 1897--1950.  This confirmed the suggested period as 5.786038$\pm$0.000078 days just after the 1919 nova event, measured the $\dot{P}$ to be ($+$1.1$\pm$0.4)$\times$10$^{-8}$ from 1923--2015, and measured the orbital period change across the 1919 eruption of $-$0.001587$\pm$0.000284 days.  The post-eruption $\dot{P}$ can be updated and improved with the 24.4 days of $TESS$ photometry for Sector 13 (middle 2019) with 1800 s exposures.  This shows a prominent ellipsoidal modulation of 22 per cent full-amplitude on average with a period near 2.89 d, showing no irradiation effects or eclipses, although orbit-to-orbit variations are apparent in both minimum and maximum fluxes as well as the peak-to-peak time intervals.  I fit a simple sinewave to the $TESS$ light curve, and derive a time of minimum to be HJD 2458673.9937$\pm$0.0092.  With this added time of minimum, I now derive a $\dot{P}$ equal to ($+$9.5$\pm$2.9)$\times$10$^{-9}$.  This update makes for a small change in $\Delta$P, which is now $-$0.00155$\pm$0.00028 days, which is to say that the orbital period and orbital radius both {\it decreased} across the eruption.

{\bf U Sco}  is a recurrent nova with 11 observed eruptions from 1863 to 2022, plus one eruption missed in a solar gap peaking at 2016.78$\pm$0.10 (Schaefer 2022a).  I discovered deep eclipses back in 1990 (Schaefer 1990), and have been intensely collecting eclipse times ever since (Schaefer  2011; 2022a; Schaefer et al. 2011; Schaefer \& Ringwald 1995), as the cornerstone of my long-term program of measuring period changes in classical and recurrent novae (e.g., Schaefer 2020b).  Schaefer (2022a) has measured $\dot{P}$ during four inter-eruption intervals, and measured $\Delta P$ across three eruptions.  My original motivation was to measure the $\Delta P$ to get $M_{ejecta}$ to test whether RNe can be progenitors of Type {\rm I}a supernova.  Disappointingly, I found that measured $\Delta P$ values are being dominated by unknown physical mechanisms (Schaefer 2020b), so $M_{ejecta}$ cannot be derived with useable confidence.  Instead, I found that the period changes in U Sco are mysterious and vary greatly from eruption-to-eruption.  The $\dot{P}$ values change by over a factor of 10$\times$ across eruptions, which is enigmatic because all the eruptions are identical photometrically and spectroscopically, so whatever physical mechanism during eruption is not manifesting in a visible manner.  The $\Delta P$ values also change by more than one order-of-magnitude from eruption-to-eruption, again with no photometric or spectroscopic manifestation.  Further, the large $\Delta P$ values cannot be explained by standard models, because the $M_{ejecta}$ must certainly be too small to produce the large sudden changes in $P$.  With the last eruption peaking in June 2022 and coming to completion after 60 days, there has been little time to get post-eruption eclipse timings.  The timings during the tail of the eclipse are skewed from the broken parabola model due to the centre of light shifting from quiescence to eruption, as was well observed in prior eruptions (Schaefer 2011).  As an update, G. Myers (AAVSO) has three post-eruption eclipse timings at HJD 2459808.9980$\pm$0.0027, 2460034.1978$\pm$0.0006, and 2460039.1231$\pm$0.0024.  These now cover too short an interval to derive the $\Delta P$ or the new $\dot{P}$.

\section{TWELVE $\Delta$P AND TWENTY $\dot{P}$ MEASURES} 

\begin{table*}
	\centering
	\caption{Measured and theoretical values for $\Delta P$/$P$ for 14 nova eruptions}
	\begin{tabular}{lllcllllll} 
		\hline
		Nova &    $P$ &      Eruption   &  $\frac{\Delta P}{P}$ observed   &   $\frac{\Delta P_{\rm ml}}{P}$   &   $\frac{\Delta P_{\rm FAML}}{P}$    &   $\frac{|\Delta P_{\rm jet}|}{P}$     &     $\frac{\Delta P_{\rm HibM}}{P}$      \\
		  &  days &      Year   &  ppm   &   ppm   &   ppm    &  ppm     &     ppm      \\
		\hline	
		
CI Aql	&	0.618	&	2000	&	2.5	$\pm$	1.9	&	0.97	&	-0.0089	&	$<$	9	&	$>$	3622	\\
QZ Aur	&	0.358	&	1964	&	-290.31	$\pm$	0.20	&	10.47	&	-0.21	&	$<$	42	&	$>$	1575	\\
T CrB	&	227	&	1946	&	815	$\pm$	247	&	0.09	&	-0.000059	&	$<$	12	&	$>$	71839	\\
HR Del	&	0.214	&	1967	&	-472.5	$\pm$	3.4	&	163.9	&	-7.0	&	$<$	312	&	$>$	1078	\\
DQ Her	&	0.194	&	1935	&	-4.45	$\pm$	0.03	&	200.0	&	-6.1	&	$<$	549	&	$>$	1179	\\
BT Mon	&	0.334	&	1939	&	39.61	$\pm$	0.39	&	20.94	&	-0.22	&	$<$	161	&	$>$	1869	\\
RR Pic	&	0.145	&	1925	&	-2004.0	$\pm$	0.9	&	29.63	&	-1.30	&	$<$	61	&	$>$	975	\\
T Pyx	&	0.0762	&	2011	&	50.3	$\pm$	7.9	&	8.00	&	-0.11	&	$<$	65	&	$>$	891	\\
V1017 Sgr	&	5.79	&	1919	&	-268	$\pm$	48	&	6.67	&	-0.065	&	$<$	57	&	$>$	12454	\\
U Sco	&	1.23	&	1999	&	3.9	$\pm$	6.2	&	0.85	&	-0.0025	&	$<$	24	&	$>$	3586	\\
U Sco	&	1.23	&	2010	&	22.44	$\pm$	0.98	&	0.85	&	-0.0025	&	$<$	24	&	$>$	3586	\\
U Sco	&	1.23	&	2016	&	35.4	$\pm$	7.1	&	0.85	&	-0.0025	&	$<$	24	&	$>$	3586	\\

		\hline
	\end{tabular}
\end{table*}

\begin{figure}
	\includegraphics[width=1.0\columnwidth]{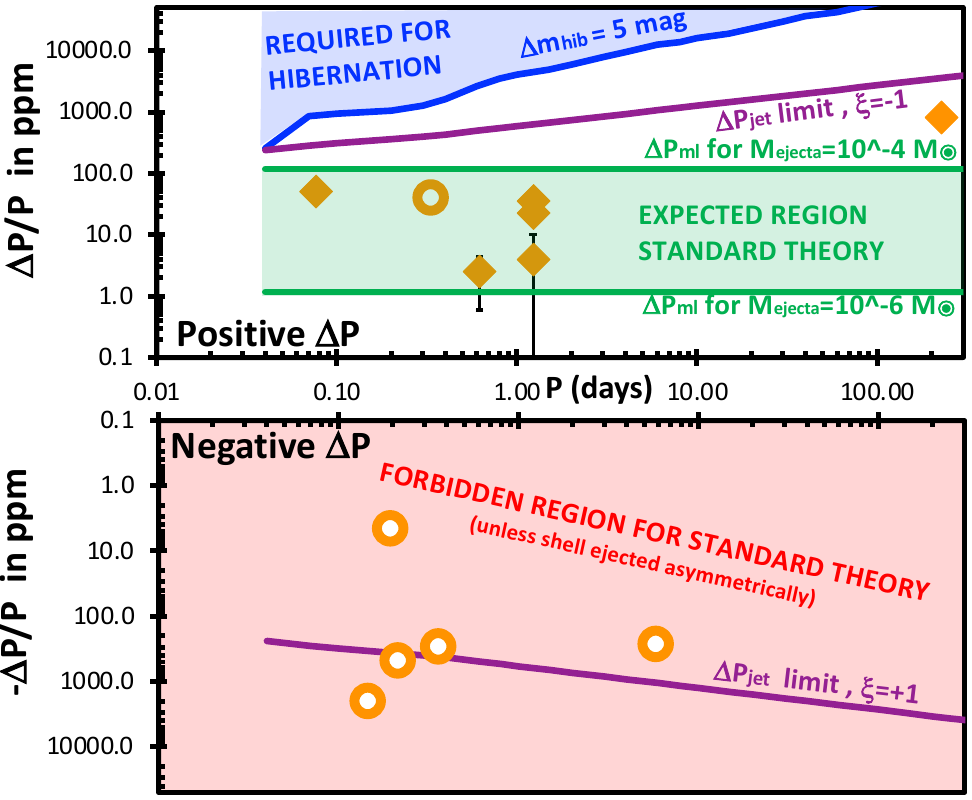}
    \caption{$\Delta P$/$P$ versus $P$ for all 12 nova eruptions.  This plot is trying to show a large range for both positive and negative values, and my solution for display to to have two stacked plots, with logarithmic scales, where the top panel is for plotting {\it positive} $\Delta P$/$P$ values, and the bottom panel is for plotting {\it negative} values.  The sudden period changes for the classical nova eruptions are indicated with a round orange symbol filled with white, while the recurrent nova eruptions are indicated with the filled orange diamonds.  For unknown reasons, all the RNe have positive $\Delta P$/$P$, while all the negative values are for CNe.  The green shaded region is bounded by the standard theory predictions for the cases of $M_{\rm ejecta}$ equal to 10$^{-6}$ and 10$^{-4}$ $M_{\odot}$, with this spanning the expected cases.  Critically, standard theory cannot explain any case where the $\Delta P$/$P$ is negative (as depicted by the red-shaded forbidden region covering the entire lower panel).  This alone proves that some additional physical mechanism outside the standard model must exist and dominate the majority of the cases.  With this the standard model has failed its strong predictions.  Further, the Hibernation Model requires that the nova eruption be in the blue-shaded region at the top of the upper plot so as to produce a post-eruption drop in in magnitude that is large enough to be called `hibernation'.  The Hibernation Model has in all cases failed this critical prediction and requirement.  Indeed, five of the nova eruptions have {\it negative} $\Delta P$/$P$, which is to say that the binary orbit has {\it contracted}, leading to the `anti-Hibernation' case where the post-eruption steady-state magnitudes should actually be more luminous than the pre-eruption-magnitudes, the exact opposite of the HibM.  }
\end{figure}

\begin{table*}
	\centering
	\caption{Measured and theoretical values for $\dot{P}$ for 20 inter-eruption intervals}
	\begin{tabular}{lllclllll}
		\hline
		Nova &   $O-C$ coverage  &  $\Delta Y$   & $\dot{P}$ observed  &  $\dot{P}_{\rm mt}$  &  $\dot{P}_{\rm GR}$  &  $\dot{P}_{\rm mb}$  &  $\dot{P}_{\Delta P}$  &  $|\dot{P}_{\rm 3rd*}|$  \\
		 &  & years     &  10$^{-9}$  &  10$^{-9}$  &  10$^{-9}$  &  10$^{-9}$  &  10$^{-9}$  &  10$^{-9}$  \\
		\hline	
CI Aql	&	1991--2001	&	10	&	-0.47	$\pm$	0.60	&	1.27	&	-0.000039	&	-0.0073	&	0.18	&	$\ll$	7.49	\\
CI Aql	&	2001--2022	&	21	&	-0.516	$\pm$	0.045	&	1.27	&	-0.000039	&	-0.0073	&	0.18	&	$\ll$	2.79	\\
T Aur	&	1954--2021	&	67	&	-0.0054	$\pm$	0.0024	&	0.055	&	-0.000113	&	-0.0080	&	0.12	&	$\ll$	0.14	\\
QZ Aur	&	1990--2021	&	31	&	-0.039	$\pm$	0.014	&	0.18	&	-0.000089	&	-0.0122	&	-0.68	&	$\ll$	0.99	\\
V394 CrA	&	1989--2021	&	32	&	-0.5	$\pm$	0.9	&	0.41	&	-0.000010	&	-0.0056	&	0.12	&	$\ll$	3.72	\\
T CrB	&	1867--1946	&	79	&	1750	$\pm$	4500	&	236	&	0.000000	&	-0.0009	&	6331	&	$\ll$	171	\\
T CrB	&	1947--2022	&	75	&	-8900	$\pm$	1600	&	236	&	0.000000	&	-0.0009	&	6331	&	$\ll$	183	\\
V1500 Cyg	&	1978--2022	&	44	&	-0.027	$\pm$	0.010	&	0.015	&	-0.000120	&	-0.0024	&	0.000028	&	$\ll$	0.16	\\
HR Del	&	1967--2022	&	55	&	0.163	$\pm$	0.037	&	0.37	&	-0.000098	&	-0.0105	&	-0.55	&	$\ll$	0.33	\\
DQ Her	&	1954--2022	&	68	&	0.00013	$\pm$	0.00012	&	0.014	&	-0.000080	&	-0.0080	&	-0.000030	&	$\ll$	0.15	\\
BT Mon	&	1941--2020	&	79	&	-0.068	$\pm$	0.003	&	0.0055	&	-0.000099	&	-0.0107	&	0.0012	&	$\ll$	0.15	\\
IM Nor	&	2003--2021	&	18	&	0.0251	$\pm$	0.0007	&	0.17	&	-0.000208	&	-0.0026	&	0.0049	&	$\ll$	0.67	\\
RR Pic	&	1945--2021	&	76	&	0.0958	$\pm$	0.0034	&	0.42	&	-0.000187	&	-0.0059	&	-0.80	&	$\ll$	0.14	\\
T Pyx	&	1986--2011	&	25	&	0.649	$\pm$	0.007	&	0.72	&	-0.000361	&	-0.0027	&	0.44	&	$\ll$	0.31	\\
T Pyx	&	2011--2022	&	11	&	0.367	$\pm$	0.027	&	0.72	&	-0.000361	&	-0.0027	&	0.44	&	$\ll$	0.93	\\
V1017 Sgr	&	1919--2019	&	100	&	9.5	$\pm$	2.9	&	0.44	&	-0.000001	&	-0.0030	&	-1.41	&	$\ll$	1.97	\\
U Sco	&	1987--1992	&	5	&	-3.2	$\pm$	1.9	&	1.58	&	-0.000016	&	-0.0064	&	6.73	&	$\ll$	19.7	\\
U Sco	&	1999--2010	&	11	&	-1.1	$\pm$	1.1	&	1.58	&	-0.000016	&	-0.0064	&	6.73	&	$\ll$	6.87	\\
U Sco	&	2010--2016	&	6	&	-21.1	$\pm$	3.2	&	1.58	&	-0.000016	&	-0.0064	&	6.73	&	$\ll$	15.4	\\
U Sco	&	2016--2022	&	6	&	-8.8	$\pm$	2.9	&	1.58	&	-0.000016	&	-0.0064	&	6.73	&	$\ll$	15.4	\\

		\hline
	\end{tabular}
\end{table*}

\begin{figure}
	\includegraphics[width=1.0\columnwidth]{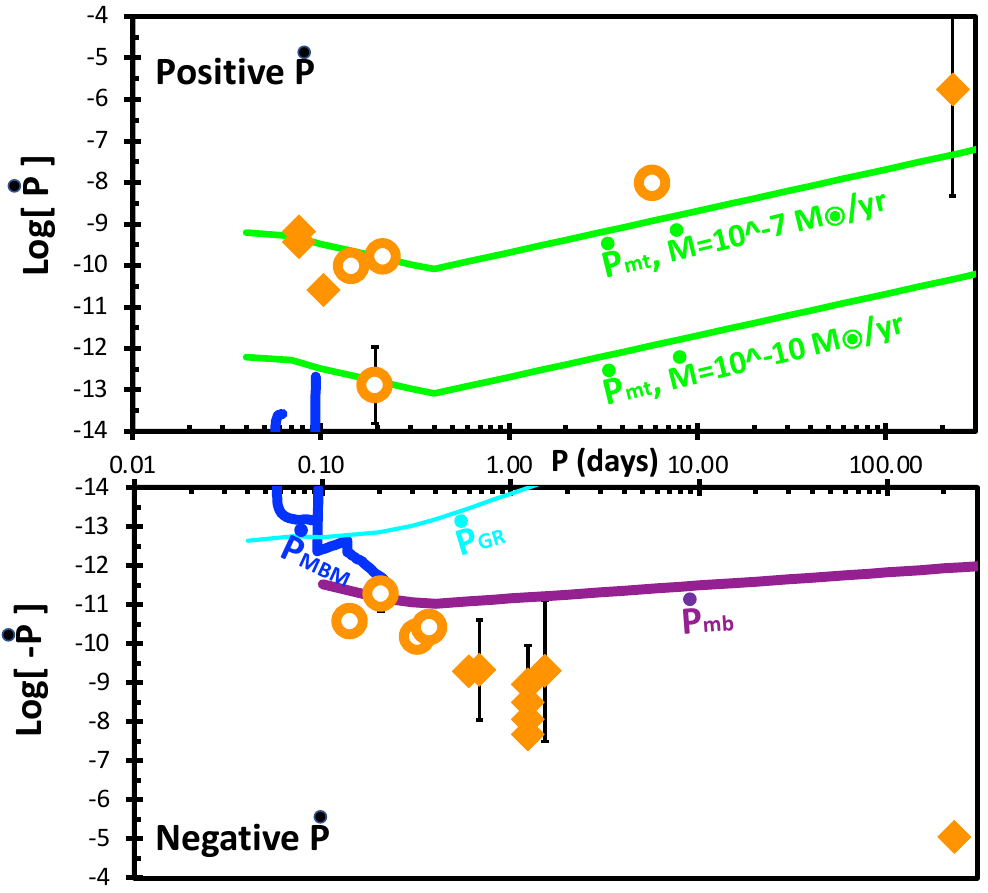}
    \caption{$\dot{P}$ versus $P$ for all 20 inter-eruption intervals.  This plot is trying to show a large range for both positive and negative values, and my solution for display to to have two stacked plots, with logarithmic scales, where the top panel is for plotting {\it positive} $\dot{P}$ values, and the bottom panel is for plotting {\it negative} values.  (The two diamonds for CI Aql are slightly offset in $P$ to avoid overlap hiding one point.)  The predicted evolutionary track of the Magnetic Braking Model is shown as a blue curve in the upper-left of the lower panel.  The basic mechanism of mass transfer can only make for a positive $\dot{P}$, with $\dot{P}_{mt}$ represented by the two green curves in the upper panel, with the curves for mass transfer rates of 10$^{-7}$ and 10$^{-10}$ M$_{\odot}$ yr$^{-1}$.  The effects of the gravitational wave emission, $\dot{P}_{GR}$, are shown by the cyan curve in the upper-left of the lower panel.  The effects of magnetic braking, $\dot{P}_{mb}$, are always negative, and are depicted as a curve across the upper part of the lower panel.  A primary science point of this plot is to show the huge spread, both positive and negative.  Pointedly, most novae have the observed $\dot{P}$ disagreeing with the MBM predictions by large factors.  The existence of this large scatter forces the recognition that the MBM is missing some unknown physical mechanism whose effects dominate over the modelled effects by orders of magnitude. }
\end{figure}

\begin{table*}
	\centering
	\caption{Properties of the 14 Novae}
	\begin{tabular}{llllllrrrlr} 
		\hline
		Nova &  Eruption classes  & $P$ & $M_{\rm WD}$  &  $M_{\rm comp}$   &   $M_{\rm ejecta}$   &   $V_{\rm ejecta}$    &    $T_{\rm comp}$     &     $M_{\rm V,q}$     &     $\dot{M}$     &     $\tau_{\rm rec}$  \\
		 &  &   days & $M_{\odot}$  &  $M_{\odot}$   &   $M_{\odot}$   &   km s$^{-1}$    &    K     &     mag    &     $M_{\odot}$ yr$^{-1}$     &     year  \\
		\hline	
CI Aql	&	RN, P(32)	&	0.618	&	1.21	&	0.85	&	1$\times$10$^{-6}$	&	2250	&	7600	&	1.3	&	1$\times$10$^{-7}$	&	24	\\
T Aur	&	CN, D(84), Fe {\rm II}, shell	&	0.204	&	0.80	&	0.50	&	1$\times$10$^{-4}$	&	1100	&	3700	&	4.0	&	1$\times$10$^{-8}$	&	700	\\
QZ Aur	&	CN	&	0.358	&	0.98	&	0.93	&	1$\times$10$^{-5}$	&	1000	&	5000	&	3.2	&	3$\times$10$^{-8}$	&	420	\\
V394 CrA	&	RN, P(5)	&	1.52	&	1.34	&	0.94	&	1$\times$10$^{-6}$	&	4600	&	5500	&	3.4	&	2$\times$10$^{-8}$	&	30	\\
T CrB	&	RN, S(6)	&	227	&	1.35	&	0.81	&	1$\times$10$^{-7}$	&	4980	&	2870	&	0.0	&	6$\times$10$^{-8}$	&	80	\\
V1500 Cyg	&	CN, S(4), Hybrid \& Neon, shell	&	0.140	&	1.15	&	0.2	&	5$\times$10$^{-6}$	&	3600	&	3300	&	5.6	&	2$\times$10$^{-9}$	&	100000	\\
HR Del	&	CN, J(231), Fe {\rm II} \& Neon, shell	&	0.214	&	0.67	&	0.55	&	1$\times$10$^{-4}$	&	525	&	3700	&	2.0	&	6$\times$10$^{-8}$	&	500	\\
DQ Her	&	CN, D(100), Fe {\rm II} \& Neon, shell	&	0.194	&	0.60	&	0.40	&	1$\times$10$^{-4}$	&	800	&	3600	&	5.7	&	2$\times$10$^{-9}$	&	80000	\\
BT Mon	&	CN, F(182), shell	&	0.334	&	1.04	&	0.87	&	2$\times$10$^{-5}$	&	2100	&	6000	&	6.6	&	9$\times$10$^{-10}$	&	30000	\\
IM Nor	&	RN, P(80)	&	0.103	&	1.21	&	0.20	&	1$\times$10$^{-6}$	&	2380	&	3300	&	2.6	&	3$\times$10$^{-8}$	&	82	\\
RR Pic	&	CN, J(122), Fe {\rm II}, shell	&	0.145	&	0.95	&	0.40	&	2$\times$10$^{-5}$	&	850	&	3400	&	3.5	&	1$\times$10$^{-7}$	&	1000	\\
T Pyx	&	RN, P(22), Hybrid, shell	&	0.0762	&	1.30	&	0.2	&	6$\times$10$^{-6}$	&	5350	&	3000	&	2.0	&	2$\times$10$^{-7}$	&	24	\\
V1017 Sgr	&	CN \& DN	&	5.79	&	1.10	&	0.7	&	6$\times$10$^{-6}$	&	1000	&	5200	&	2.1	&	4$\times$10$^{-9}$	&	3000	\\
U Sco	&	RN, PP(3), He/N	&	1.23	&	1.36	&	1.0	&	1$\times$10$^{-6}$	&	5700	&	5300	&	3.4	&	9$\times$10$^{-8}$	&	10.3	\\

		\hline
	\end{tabular}
\end{table*}

My measured period changes for these 14 novae includes 12 $\Delta P$ values for 10 novae, plus 20 $\dot{P}$ values for all 14 novae.  The observed values of $\Delta P$ have been collected in Table 8.  It is convenient to express and compare the $\Delta P$/$P$ values as dimensionless numbers in  parts-per-million (ppm).  The observed values of $\dot{P}$ have been collected in Table 9.  For comparison with values in prior papers and with theory calculations, it is important to know that the $\dot{P}$values in this paper are all in dimensionless units (i.e., days/day or seconds/second) where the quadratic term in the $O-C$ parabolas is 0.5$P$$\dot{P}$$N^2$.  

Importantly, these eclipse timings and the measured $\dot{P}$ values are not associated with any transient variations associated with the nova eruption.  {\it Small} variations in $O-C$ are seen for V1500 Cyg, U Sco, and YZ Ret {\it during} the eruption (Patterson 1979; Schaefer et al. 2011; Schaefer 2022b).  These variations have vanished by the time the nova brightnesses return to their quiescent levels.  All of the eclipse times used are for years after the end of the eruption, and most are from decades to over a century after the end of the eruption.

I have plotted $\Delta P$/$P$ versus $P$ in Figure 9, and plotted $\dot{P}$/$P$ versus $P$ in Figure 10.  In both of these figures, I need to plot a wide range of values, both positive and negative, and this makes for a problem of presentation.  My solution to this presentation problem is to stack two plots with the upper plot having a log-scale for the positive values, with the lower plot having a log scale where the negative values are plotted as the log of the absolute value.  The horizontal axis will be a log-scale for $P$, so as to cover the entire range from 0.0762 days to 227 days, with the scales being identical for both the upper and lower panels.  The region between the panels (with the axis labels for $P$) corresponds to values close to zero.

For making specific numerical predictions for these 14 novae, we need to have measures or estimates of various system properties, as collected in Table 10.  These input values come from a wide selection of sources.  The eruption classes for recurrence (i.e., CN versus RN), light curve classes (see Strope, Schaefer, \& Henden 2010), spectral classes, and whether an expanding nova shell has been seen are all taken from the comprehensive compilation of Schaefer (2022c).  The $P$ are all taken from my review article on orbital periods (Schaefer 2022b).  The $M_{\rm WD}$ values are taken from the measures reported in Ritter \& Kolb (2003) as well as the model values of Hachisu \& Kato (2019) and Shara et al. (2018).  The $M_{\rm comp}$ values are taken from studies of the individual novae (Schaefer 2020b; 2023; Schaefer et al. 2013; Salazar et al. 2017) as well as from the model of Knigge et al. (2011) as based only on the orbital period.  The mass of the ejecta only has badly-measured values (see Appendix A of Schaefer 2011), so I have been forced to use estimates from individual studies (Schaefer 2020b) and from general models (Yaron et al. 2005) that start with accretion rates and trigger masses, so the tabulated $M_{\rm ejecta}$ are only crude approximations.  The $V_{\rm ejecta}$ values are mostly from measured widths of emission lines (as compiled by Schaefer 2022c and Payne-Gaposchkin 1964).  QZ Aur has no reported velocity, so I have adopted typical values for its light curve class for its $M_{\rm WD}$ and $\dot{M}$.  These velocities are poorly defined, as the reported values are variously FWHM, HWHM, and HWZI of some hydrogen emission line, while the velocities change substantially from line-to-line and and over time around the peak of the nova event, so my tabulated values only have an accuracy and consistency of perhaps a factor-of-two.  The $T_{\rm comp}$ values come from blackbody fits to spectral energy distributions (Schaefer 2010; 2023; Salazar et al. 2017), from models keyed to $P$ (Knigge et al. 2011), and from positions along evolutionary tracks for the estimated size and mass of the companion stars.  The $M_{\rm V,q}$ values are calculated from the distances and $E(B-V)$ values in Schaefer (2022c) plus the quiescent $V$ magnitudes from Schaefer (2010), Schaefer et al. (2019), Salazar et al. (2017), and Strope et al. (2010).  The accretion rates are estimated from individual studies (Schaefer 2022a; 2023; Schaefer et al. 2013) and from $M_{\rm V,q}$ (with corrections to subtract the companion star flux) by an approximate relation based on Dubus, Otulakowska-Hypka, \& Lasota (2018).  The recurrence times are taken from Schaefer (2010) for the RNe, while the others have estimated $\tau_{\rm rec}$ values from the models of Yaron et al. (2005) as based on their $M_{\rm WD}$ and $\dot{M}$.  In all, many of the tabulated properties in Table 10 are only approximate values with poor accuracy.  Fortunately, for the purposes at hand of making model predictions of period changes, the comparisons to the observed values will be order-of-magnitude, so the uncertainties in the input will be of little import.

The theoretical predictions need six further sets of values.  These are the radius of the companion star $R_{\rm comp}$ in units of solar-radius, the semi-major axis of the binary system $a$ in units of solar-radius, the orbital velocity of the companion star $V_{comp}$ in units of kilometers per second, the orbital velocity of the white dwarf $V_{\rm WD}$ in units of kilometers per second, the companion star's atmospheric scale height at the surface of the Roche lobe $H$ in units of kilometers, and the dimensionless mass ratio $q$ which equals $M_{\rm comp}$/$M_{\rm WD}$.  These quantities are all directly calculated from the input in Table 10.  The equations for these calculations are all standard and well-known (e.g., in chapter 4 of Frank, King, \& Raine 2002).

\section{TESTING THE HIBERNATION MODEL AND STANDARD THEORY WITH THE OBSERVED $\Delta P$}

\subsection{Standard Theory for $\Delta P$ mechanisms}

Only three mechanisms for $\Delta P$ are well-known, derived in detail, and accepted in the literature.  (See Schaefer 2020a and Martin et al. 2011 for reviews.)  

{\bf (1)} The first mechanism is the mass loss (with subscript `ml'), when the sudden loss of $M_{\rm ejecta}$ changes the $P$ by Kepler's Law.  Mass loss always leads to a period increase, a positive $\Delta P$, and a larger orbital separation.  This first mechanism has $\Delta P_{\rm ml}$ dominating greatly over the other two mechanisms.   The period change due to a sudden mass loss in the system is 
\begin{equation}
\Delta P_{\rm ml} = A P \frac{M_{\rm ejecta}}{M_{\rm WD}}.
\end{equation}
$A$ is a complex factor collecting some small effects, as given in equations 4 and 5 of Schaefer (2020a).  In all cases, $A$ is close to $2/(1+q)$.  Importantly, $\Delta P_{\rm ml}$/$P$ is always positive.  With the mass loss mechanism dominating over the other two mechanisms, this means that the total predicted period change from the standard theory cannot be negative.  The forbidden region with negative-$\Delta P$ is depicted in Fig. 9 with the red-shading over the entire lower panel.  The range of $\Delta P_{\rm ml}$/$P$ depends primarily on $M_{\rm ejecta}$, which typically varies from 10$^{-6}$ and 10$^{-4}$ $M_{\odot}$.  Fig. 9 shows horizontal green lines for these two cases.  The green-shaded region between the two green lines is where standard theory expects the position for most novae.

{\bf (2)} The second mechanism is called `frictional angular momentum loss' (`FAML'), when the companion star travels through the expanding shell of the nova event with dynamical effects slowing its forward velocity making for a short interval of $P$ decreasing.  $\Delta P_{\rm FAML}$ is always negative, and is always greatly smaller than $\Delta P_{\rm ml}$. The period change due to FAML is 
\begin{equation}
\Delta P_{\rm FAML} = -0.75 P \frac{M_{\rm ejecta}}{M_{\rm comp}} \frac{V_{\rm comp}}{V_{\rm ejecta}} \frac{R_{\rm comp}}{a}.
\end{equation}
This mechanism can only produce a negative value, which is to say that the period decreases and the binary orbit gets suddenly tighter.  For the factors in equation 3, $V_{\rm comp}/V_{\rm ejecta}$$\ll$1 and $R_{\rm comp}/a$$<$0.5.  In comparing the equations for mass loss and FAML, $\Delta P_{\rm FAML}$ will always be greatly smaller than $\Delta P_{\rm ml}$, so the sum of the two effects will always be positive and close to $\Delta P_{\rm ml}$.

{\bf (3)} The third mechanism can be described as magnetic braking in the nova shell by the companion star (with subscript `mbns'), where a hypothetical substantial magnetic field on the companion will entrain outflowing ejecta to co-rotate, changing the spin of the companion, with this angular momentum change rapidly going into the orbit by the usual tidal effects.  The period change due to magnetic braking of the companion star inside the expanding nova shell is 
\begin{equation}
\frac{\Delta P_{mbns}}{P} =  -0.75 \frac{M_{\rm ejecta}}{M_{\rm WD}} \left[  \left(   \frac{1-q}{q} \frac{R_{\rm A}^2}{a^2} \right) +  \left(  \frac{1+q}{q} \frac{R_A^4}{a^4} \right) \right].
\end{equation}
Here, $R_A$ is the Alfv\'{e}n radius for the companion star's magnetic field, which is proportional to the one-third power of the field's magnetic moment.  In comparing this with Eq. 2, we see that the magnetic braking effect is negligibly small except when $R_A/a$ is near unity and $q$ is small.  Even with $R_A/a$=1 and a very high $M_{\rm ejecta}$=10$^{-4}$ M$_{\odot}$, a typical nova with $P$=0.4 days will have $\Delta P_{mbns}$/$P$=$-$16 ppm, with this being 7$\times$ smaller than $\Delta P_{ml}$/$P$.  The magnetic field required to get $R_A/a$=1 is 0.22 megaGauss at the surface of a typical nova companion star, and this is unreasonably large.  For all realistic cases, $\Delta P_{mbns}$ is three orders of magnitude or more smaller than $\Delta P_{ml}$.  The total period change from standard theory will be $\Delta P_{\rm ml}$+$\Delta P_{\rm FAML}$+$\Delta P_{\rm mbns}$, and this will always be close to $\Delta P_{\rm ml}$.

\subsection{Tests of the standard theory for $\Delta P$ mechanisms}


Now we can make the critical test of standard theory by comparing the observed and predicted period changes, with the evaluations summarized in Table 11.  There is reasonable agreement for the two cases of CI Aql and BT Mon.  However, there is stark contradictions for the remaining ten eruptions.  Five of the novae show negative $\Delta P$, which is impossible by the standard theory.  T Pyx has a well-measured $\Delta P$ that is near 6$\times$ larger than the theory prediction, and there is no chance\footnote{This can be seen from the models of Yaron et al. (2005) and Shen \& Bildsten (2009).  Further, to accrete this mass over the preceding 44 years would require an average accretion rate of 8$\times$10$^{-7}$ M$_{\odot}$ yr$^{-1}$, with this being impossible as lying above the stable hydrogen burning regime.} that its 2011 eruption had $M_{\rm ejecta}$ as large as 3.6$\times$10$^{-5}$ M$_{\odot}$.  T CrB does not fit into the standard theory, as its predicted period change is 8800$\times$ smaller than observed.  So in most cases, the standard models for the $\Delta P$ mechanisms have strongly failed in their predictions.

The standard model also fails badly in not being able to explain the large changes in $\Delta P$ for the three eruptions of U Sco.  In particular, the only way to change $\Delta P$ from theory for a given nova system is to vary $M_{\rm ejecta}$, whereas U Sco certainly does not have order-of-magnitude changes in $M_{\rm ejecta}$.  The strong observational conclusion from many eruptions is that all the U Sco eruptions are identical both photometrically and spectroscopically (Schaefer 2010), and such can only occur when all the eruptions deliver the same ejecta mass.  So the standard theory for $\Delta P$ requires an essentially constant value from eruption-to-eruption, and this is not what is observed in the only case of record.  By itself, this variability of $\Delta P$ serves as a refutation for the standard theory, for at least this one case of U Sco.

A possibly serious problem for the standard nova theory comes from the measure that the break in the $O-C$ curve for T Pyx was 108.4$\pm$12.6 days after the start of the eruption.  For the standard mechanisms, the effects on $P$ will be effectively simultaneous with the ejection, with the rate of change being proportional to the ejection rate.  The observed $O-C$ curve break means that half of $M_{\rm ejecta}$ was blown off the white dwarf {\it after} day 108.4 of the eruption.  If the mass ejection is constant up until some turnoff time, then this schematic picture can produce a break at 108.4 days only if ejection continues full force up until day 216.8.  This seems implausible because on day 216.8, T Pyx is near 6 mags below peak.  For a schematic model where the ejection rate is proportional to the flux level above quiescence, I have integrated the eruption $V$ light curve from AAVSO to see that 50 per cent ejection is at an epoch of 39 days after the start of the eruption, and with 95 per cent ejection by day 108.4.  Chomiuk et al. (2014)  has modelled the T Pyx ejection as being two-staged with an initial ejection on day 0 and a second ejection on day $\sim$65.  These ejection time-profiles are greatly inconsistent with the observed middle-ejection date of 108.4 days after the eruption's beginning.  On the face of it, I could declare another failed prediction of nova models.  However, I am reluctant to make such a stark declaration because models of nova mass ejection are apparently with sufficient uncertainty such that substantial late-ejection might be possible.  Nevertheless, standard nova theory has a serious problem, which can now only be solved by a detailed physical model demonstrating that T Pyx ejected half its mass after day 108.4.

The standard theory has failed all the tests of its prediction.  And these are critical direct tests of the fundamental properties, all for many nova eruptions on a wide array of nova systems.  So the easy conclusion is that the standard theory is wrong and not useful.  This is a rather stark conclusion that is strongly pointed to by the evidence.  Nevertheless, the physical models for the mechanisms are undoubtedly correct.  So what is going on?  The only reconciliation that I can think of is that there must be some additional mechanism that has not been included into the three mechanisms of standard theory.  That is, we need some now-unrecognized fourth mechanism to add in with the three known mechanisms.  This fourth mechanism must be large, usually substantially larger than the simple effects of mass loss ejected by the white dwarf.  This fourth mechanism will dominate over the currently accepted theory.

\begin{table}
	\centering
	\caption{Agreement between observed $\Delta P$ and $\dot{P}$ values and theory}
	\begin{tabular}{llll} 
		\hline
		Nova &  $\Delta P$ in  & $\Delta P$ in  &  $\dot{P}$ in MBM and \\
		 &  standard theory  &   Hibernation & standard theory  \\
		\hline	
CI Aql	&	\large \checkmark	&	$\mathbf{X}$ (5)	&	$\mathbf{X}$ (9)	\\
T Aur	&	...	&	...	&	\large \checkmark	\\
QZ Aur	&	$\mathbf{X}$ (1)	&	$\mathbf{X}$ (5), $\mathbf{X}$ (6)	&	\large \checkmark	\\
V394 CrA	&	...	&	...	&	...	\\
T CrB	&	$\mathbf{X}$ (4)	&	$\mathbf{X}$ (5)	&	$\mathbf{X}$ (9), $\mathbf{X}$ (10)	\\
V1500 Cyg	&	...	&	...	&	$\mathbf{X}$ (9)	\\
HR Del	&	$\mathbf{X}$ (1)	&	$\mathbf{X}$ (5), $\mathbf{X}$ (6)	&	$\mathbf{X}$ (7)	\\
DQ Her	&	$\mathbf{X}$ (1)	&	$\mathbf{X}$ (5), $\mathbf{X}$ (6)	&	$\mathbf{X}$ (7)	\\
BT Mon	&	\large \checkmark	&	$\mathbf{X}$ (5)	&	\large \checkmark	\\
IM Nor	&	...	&	...	&	{\large \checkmark} (8)	\\
RR Pic	&	$\mathbf{X}$ (1)	&	$\mathbf{X}$ (5), $\mathbf{X}$ (6)	&	$\mathbf{X}$ (7)	\\
T Pyx	&	$\mathbf{X}$ (4), {\large \textbf{?}} (3)	&	$\mathbf{X}$ (5)	&	$\mathbf{X}$ (10), {\large \checkmark} (8)	\\
V1017 Sgr	&	$\mathbf{X}$ (1)	&	$\mathbf{X}$ (5), $\mathbf{X}$ (6)	&	$\mathbf{X}$ (7)	\\
U Sco	&	$\mathbf{X}$ (2)	&	$\mathbf{X}$ (5)	&	$\mathbf{X}$ (9), $\mathbf{X}$ (10)	\\

		\hline
	\end{tabular}

\begin{flushleft}	
Notes: {\bf{(1)}} The observed $\Delta P$ is negative, a case that is forbidden by standard theory, although it might possible if the nova shell is ejected with a substantial asymmetry.
{\bf{(2)}} The observed $\Delta P$ values are substantially and significantly different from eruption-to-eruption, and such is impossible in the standard theory for nova eruptions that are all identical spectroscopically and photometrically.
{\bf{(3)}} The break in the $O-C$ curve (i.e., $\Delta P$) occurs over a small time interval centred on 108.4$\pm$12.6 days after the start of the eruption.  The problem for standard theory is that this is long after most of the mass has already been ejected.
{\bf{(4)}} The observed $\Delta P$ is greatly and significantly larger than is possible with standard theory.
{\bf{(5)}} The observed $\dot{P}$ is greatly smaller than required by HibM, so no hibernation is occurring.
{\bf{(6)}} The observed $\dot{P}$ is negative, so the orbital period is decreasing, and the binary stars are getting closer together, in a case of `anti-hibernation'.
{\bf{(7)}} The observed $\dot{P}$ is positive, and this is impossible within MBM, and is impossible within standard theory (unless the $\dot{M}$ is very large).
{\bf{(8)}} The observed $\dot{P}$ is positive, with this being impossible in the MBM of Knigge et al. (2011), but it is possible within standard theory for the case where $\dot{M}$ is very high.  For T Pyx and IM Nor, the MBM has not allowed for the possibility that these systems can have century-long intervals of accretion near the maximum (hence leading to the series of RN events) despite the period in-or-below the Period Gap.
{\bf{(9)}} The observed $\dot{P}$ is greatly more negative than possible within MBM or standard theory.
{\bf{(10)}} The observed $\dot{P}$ changed greatly at the time of a nova event, and such is not possible within MBM or standard theory.

\end{flushleft}

\end{table}

\subsection{Tests of the Hibernation Model}

The Hibernation Model has the cause of the hibernation state being that the nova eruption makes for a period {\it increase} so that the binary separation increases making for a dramatic drop in the accretion rate.  The hibernation state is driven by the period increase.  Now, from Table 8, we see that five-out-of-six classical novae have highly significant period {\it decreases}.  These negative $\Delta P$ cases can be labelled as `anti-hibernation'.  This proves that most CNe do not have the hibernation mechanism in operation.  This is a refutation of the HibM. 

HibM makes a strong prediction that the measured $\Delta P$ values be sufficiently large to drop the accretion rate by a large amount.  For the hibernation state to occur, by definition, the accretion must drop to a near-zero value.  For the hibernation state to be apparent in light curve, the drop in accretion must be such that the change is substantially larger than the usual variations displayed by old novae, for which variations by 2.5 mag (effectively 10$\times$ in accretion rate) are occasionally seen.  So a formal criterion for the case of hibernation can be to require that the system brightness (which closely follows the accretion rate for most systems) decrease by 5 magnitudes (100$\times$ in accretion rate) in brightness across each nova eruption.  Schaefer (2020a) derived the magnitude drop ($\Delta m_{\rm hib}$) as 
\begin{equation}
\Delta m_{\rm hib} = 1.086 \frac{R_{\rm comp}}{H} \left [ \left (\frac{2}{3} \frac{\Delta P_{\rm hib}}{P} \right)+ \left( \frac{1}{3} \frac{M_{\rm ejecta}}{M_{\rm comp}} \right) \right].
\end{equation}
The system properties of $R_{\rm comp}$, $H$, $M_{\rm ejecta}$, and $M_{\rm comp}$ for individual novae are tabulated in Table 10, or can be calculated from the other properties.  For $\Delta m_{\rm hib}$=5, the minimum allowed period change for hibernation to occur ($\Delta P_{\rm hib}$) can be calculated.  Table 8 lists the required changes for hibernation to occur for all 12 nova eruptions.  Further, for a typical case for the companion star (with properties as a function of $P$ from Knigge et al. 2011) and with the extreme case of $M_{\rm ejecta}$=10$^{-4}$ M$_{\odot}$, I have drawn a limit curve in Fig. 9 as a blue line near the top of the upper panel.  For hibernation to occur, the nova eruption must have its period change above the blue curve, inside the blue-shaded region.  If a nova event has a period change that places it below this `hibernation required' region, then hibernation will not occur.  From Fig. 9 and Table 8, we see that all 12 nova eruptions have the observed $\Delta P$ values far below that required for hibernation.  These individual evaluations of the HibM are in Table 11.  This is a refutation of the HibM.

\subsection{The necessity for adding some new physical mechanism for sudden period changes across a nova eruption}

With the failures of the standard models, we can only look past it for further mechanisms that solve the failures.  This fourth mechanism is what provides a large decrease in $P$ so as to make the negative-$\Delta P$ measures.  But for T CrB, this extra mechanism must greatly increase the orbital period.  This extra mechanism must have large variability from eruption-to-eruption for U Sco, yet not lead to any photometric or spectroscopic changes in the eruptions.  This fourth mechanism might have a delayed reaction onto the orbit (perhaps due to some sort of a synchronization time-scale) so as to explain the late $O-C$ break by T Pyx.

A clue to the nature of this fourth mechanism comes from the correlations of the residual effect with system properties.  The effect of the fourth mechanism should be close to $\frac{\Delta P}{P}$-$\frac{\Delta P_{\rm ml}}{P}$.  In seeking correlations with the properties in Table 10, apparent correlations are with $M_{\rm ejecta}$ and $V_{\rm ejecta}$.  The largest positive effects are for low-$M_{\rm ejecta}$ and the high-$V_{\rm ejecta}$ systems, while the most negative effects are for the high-$M_{\rm ejecta}$ and the low-$V_{\rm ejecta}$ systems.  The RNe must have low ejecta masses and high ejecta velocities, so this is what creates the division of RNe into the upper panel of Fig. 9 and the CNe into the lower panel of Fig. 9.  The observed correlations have substantial scatter, but they might be a useable clue for understanding some fourth mechanism.

So a primary result of this paper is to widely force the recognition that the standard theory is incomplete and is missing the dominant effect, and to encourage investigations of new ideas and their detailed models.  It is not that the individual mechanisms have wrong models, but the standard theory for $\Delta P$ is consistently failing all its predictions.  The dominant mechanism for nova period changes is unknown, and so the general task of CV evolution is incomplete for missing the dominant effects.

\subsection{$\Delta P$ from asymmetric ejection of the nova shell}

J. Frank (Louisiana State University) has proposed a fourth mechanism (in Schaefer 2020a), wherein the ordinary nova shell ejection is asymmetric.  If the white dwarf ejects more mass in its forward direction of motion, then its orbital velocity will be slowed and the $P$ will have a sudden drop due to the jet effects of the shell ejection.  If the white dwarf ejects more mass in the backwards direction of its orbital motion, then its orbital velocity will be sped up and $P$ will suddenly increase across the eruption.  Asymmetric ejections are ubiquitous and substantial, as demonstrated by the nearly universal hemispheric concentrations of mass visible in many nova shells.  The extreme case is the fast RN in the Andromeda Galaxy, where the ejecta from all the last thousands of eruptions is all going in one persistent direction, with an apparent opening angle of near one steradian (Darnley et al. 2019, fig. 1).  So Frank's jetting mechanism must be operating in nova eruptions at some level.

Frank derived a period change for a simple case where some fraction of the shell mass is instantaneously ejected as a hemisphere in one direction.  This gives
\begin{equation}
\Delta P_{\rm jet} = -1.5 P q \xi  \left( \frac{M_{\rm ejecta}}{M_{\rm WD}+M_{\rm comp}} \right) \frac{V_{\rm ejecta}}{V_{\rm WD}}.
\end{equation}
Here, $\xi$ is a parameter describing the asymmetry of the ejecta, with +1 representing the case where all the ejecta is in a hemisphere centred on the forward direction of the white dwarf's orbital velocity, and $-$1 represents a backward facing hemisphere of ejecta.  Narrow jets will increase the range of $\xi$ to +2 to $-$2.  My crude estimates from pictures of nova shells suggests that $|\xi|$ is usually larger than 0.4 (Schaefer 2020a).  For a typical case for a CN (with $M_{\rm ejecta}$=10$^{-4}$ M$_{\odot}$ and $V_{\rm ejecta}$=1000 km/s), Fig. 9 shows purple lines for $\xi$=$+$1 in the lower panel, and for $\xi$=$-$1 in the upper panel.  These lines are roughly the limits of the effect in realistic cases.  For the specific properties of individual nova, Table 8 lists the upper limits for the $|\Delta P_{\rm jet}|$/$P$ values.

The $\Delta P_{\rm jet}$ equation assumes instantaneous ejection, or at least with ejection faster than one orbital period.  This might be true for fast novae with short $t_3$ and an S class light curve, or for novae with long orbital periods.  However, if the nova has continuous ejection over multiple orbital periods, then the effects from the forward thrust will average out half-an-orbit later with the effects of backward thrust.  A further case where $\Delta P_{\rm jet}$ is small would be if the ejection is primarily with material lifted off the surface of a common envelope around the white dwarf, presumably caused by the companion star.

This candidate fourth mechanism satisfies the needs for explaining my $\Delta P$ data.  The jetting can be either forward or backwards, so it explains why half the nova have negative $\Delta P$.  The jetting can change direction forward/backward from eruption-to-eruption, so this provides an easy explanation for how U Sco can change its $\Delta P$ value so fast.  The jetting effect can get rather large, and this explains how $|\Delta P|$ can be much larger than allowed by mass loss.  

\section{TESTING THE MAGNETIC BRAKING MODEL WITH THE OBSERVED $\dot{P}$ MEASURES}

\subsection{Standard Theory for $\dot{P}$ mechanisms}

The standard theory of CV evolution has only four physical mechanisms that produce long-term steady period changes in quiescence (Knigge et al. 2011, Schaefer 2020a):     

({\bf 1}) The first mechanism that can produce steady period changes between eruptions is the ordinary mass transfer (with subscript `mt') from the companion star to the white dwarf, wherein a simple shifting of mass in the binary makes for period changes despite the overall angular momentum of the system ($J$) remaining constant.  
\begin{equation}
\dot{P}_{mt} = 3 P (1+q) \frac{\dot{M}}{M_{\rm comp}}.
\end{equation}
If the binary has the mass donor star less massive than the accreting star, as is the case for essentially all CVs, then the $P$ will increase.  Figure 10 displays two curves for typical cases with assumed values of $\dot{M}$ of 10$^{-10}$ and 10$^{-7}$ M$_{\odot}$ yr$^{-1}$.  Table 10 presents the best estimates for $\dot{P}_{mt}$ specifically for each of the 14 novae.

({\bf 2}) The second mechanism that can produce steady period changes in quiescence is gravitational radiation (with subscript `GR'), wherein the spinning binary star is emitting weak gravitational waves that carry off angular momentum from the system.  But this mechanism is very weak, and is negligibly small for CVs above the Period Gap.  Standard theory (e.g., Eq. 19 of Rappaport, Joss, \& Webbink 1982) gives the fractional change in the binary's angular momentum ($\dot{J}$$_{GR}$/$J$).  With Kepler's Law, the conservation of angular momentum, and unchanging stellar masses, the associated period change is $\dot{P}_{GR}$=3$P$$\dot{J}_{GR}$/$J$.  With this, 
\begin{equation}
\dot{P}_{GR} = \frac{-96}{5} (2\pi)^{\frac{8}{3}} \frac{G^{\frac{5}{3}}}{c^5} M_{\rm WD} M_{\rm comp} (M_{\rm WD} + M_{\rm comp})^{\frac{-1}{3}} P^{\frac{-5}{3}}.
\end{equation}
Knigge et al. (2011) finds that they must do the equivalent of strengthening the GR effect by a factor of 2.47$\pm$0.22 so as to match the observed minimum $P$ for CVs, which is to say that their model needs to arbitrarily add angular momentum losses from some unknown mechanism for systems below the Period Gap.

 ({\bf 3})The third mechanism is labelled `magnetic braking' (with subscript `mb'), wherein a postulated stellar wind from the companion star has its outgoing material forced into co-rotation by the companion star's hypothetical magnetic field so as to carry away angular momentum from the system, which slows the rotation of the companion, which is then speedily transferred by tidal forces into a slowing of the orbital motion.  This mechanism can only make for $P$ to decrease.  Neither stellar winds nor magnetic fields on the companions stars have ever been observed, so their strengths could lie anywhere from effectively-zero up to relatively large values.  A consequence of this is that the real strength of magnetic braking is unknown by many orders-of-magnitude, and is likely greatly variable from CV-to-CV.  Further, there is no useable physics model for the effect with any confidence\footnote{The array of conflicting schematic `recipes' is shown in Appendix A of Knigge et al. (2011).  Their Fig. 2 shows `recipes' ({\it selected for their popularity and flexibility}) spanning over five orders-of-magnitude in $\dot{J}$ at a given $P$.}.  The result of this is that the size of the period changes, $\dot{P}_{mb}$, are unknown to within $>$5 orders-of-magnitude.
 
Models of CV evolution need some sort of formula to describe the unknown magnetic braking effects, so a characteristic tactic is to postulate a simple power law behaviour attributed to all individual CVs.  These typically have $\dot{J}_{mb}$$\propto$$R_{\rm comp}^{\gamma}$, where $\gamma$ is only constrained to be between 0 and 4, and the negative proportionality constant is allowed to freely vary.  Knigge et al. (2011) varied $\gamma$ and the proportionality constant so as to match the observed minimum-$P$ and the position of the Period Gap for CVs.  Further, Knigge et al. (2011) allowed for the free fitting of a constant times the GR contribution to $\dot{J}$.  With these free fittings, the agreement between the MBM and the histogram of CV periods has the three critical periods fitted by freely varying three parameters, so a close agreement is guaranteed by construction.  Thus, the MBM critical periods are not successful predictions, but rather were input by hand.

The final `recipe' for $\dot{J}_{mb}$ of Knigge et al. (2011) has the stereotypical formula from Rappaport et al. (1983, Eq. 36, with $\gamma$=3) multiplied by a factor of 0.66.  As with the GR contribution, the period change will be $\dot{P}_{mb}$=3$P$$\dot{J}_{mb}$/$J$.  With this,
\begin{equation}
\dot{P}_{mb} = -9.4\times 10^{-13} \left( \frac{P}{{\rm day}} \right)^{\frac{-7}{3}} \left( \frac{M_{\rm WD}}{M_{\odot}} \right)^{\frac{-2}{3}} \left( 1+q \right)^{\frac{1}{3}} \left( \frac{R_{\rm comp}}{R_{\odot}} \right)^3.
\end{equation}
Here, M$_{tot}$ equals M$_{\rm WD}$+M$_{\rm comp}$, and the $\dot{P}_{mb}$ is dimensionless.  This term should be zeroed out for CVs with periods below the Period Gap.  The $\dot{P}_{mb}$ is plotted as a function of $P$ with a purple curve in the lower panel of Figure 10.  Table 9 has the $\dot{P}_{mb}$ values calculated for the individual properties of the 14 novae.

The sum of the first three mechanisms, $\dot{P}_{mt} + \dot{P}_{GR} + \dot{P}_{mb}$, is the value to compare with the individual measured $\dot{P}$ that covers just one inter-eruption interval with no nova event.  With the GR effect always being small, $\dot{P}$ comes down to a competition between the positive contribution of mass transfer versus the negative contribution of magnetic braking.  In general, the magnetic braking dominates, as found for the MBM of Knigge et al. (2011), and as required for the inevitable evolution from long-period to short-period for CVs.  The only exception is for CVs with a very high accretion rate (up near the limit where steady hydrogen burning starts at $\sim$10$^{-7}$ M$_{\odot}$ yr$^{-1}$).

({\bf 4}) The fourth mechanism that can produce long term period changes arises from the sudden jerks from the nova eruptions (with subscript `$\Delta P$'), wherein the $\Delta P$ gets averaged over each eruption cycle.  With the observed $\Delta P$ averaged over the nova recurrence time $\tau_{rec}$, the effective long-term period change is
\begin{equation}
\dot{P}_{\Delta P} = \Delta P / \tau_{rec}.
\end{equation}
In the model of Knigge et al. (2011, with $\alpha$ in their Eq. 4 set equal to unity), this term is not calculated from observed values, but rather is simply $\Delta P_{ml}$/$\tau_{rec}$, with this implying a frequent very large error in their calculated $\dot{P}_{\Delta P}$ (see Table 8).  Fig. 10 displays two lines for average cases with $\tau_{rec}$=1000 years and $\Delta P$/$P$=$\pm$100 ppm.  (These lines do not account for $\tau_{rec}$ changing with $P$, so they are only plotted to set an approximate scale.)  Best estimate values for all 14 novae are presented in Table 9.

The standard theory prediction for long-term evolution calculations has the period change equalling the sum of all four effects, 
\begin{equation}
\dot{P}_{total} = \dot{P}_{mt} + \dot{P}_{GR} + \dot{P}_{mb} + \dot{P}_{\Delta P}.
\end{equation}
The MBM model of Knigge et al. (2011) has added together the same four effects to get$\dot{P}_{MBM}$, as tabulated in their Table 4, and shown as the blue curve in Fig. 10.

\subsection{Can $\dot{P}$ be an artefact of a third-body in orbit?}

The CV binary (with the white dwarf and its companion) can possibly have a third body in a wide orbit.  This ordinary orbit will have the inner binary moving closer and farther from Earth periodically, with the delays from the light travel time making for the CV eclipse times to vary with a sinewave in the $O-C$ curve.  If the eclipse times are collected over an interval of time $\Delta Y$ (the duration of observations as in the second column of Table 9) that is roughly centred on the conjunction of the third body in its orbit, then the observed part of the sinewave can look like a parabola with a measured $\dot{P}_{3rd*}$.  If the observing interval $\Delta Y$ is roughly centred around the time of superior conjunction of the third body, then the apparent $\dot{P}_{3rd*}$ will be negative.  If centred on inferior conjunction, then the $O-C$ curve will look like a concave-up parabola.  To look like a parabola over time interval $\Delta Y$, the orbital period of the third star, $P_{3rd*}$ will have to be $>$2$\Delta Y$ or so.  I know of no confirmed case where a CV has a third body in orbit, yet the frequency of wide triple systems suggests that at least some CVs have tertiary companions.  In this case, some of the novae in Table 8 might have the measured $\dot{P}$ arising from third-body orbits rather than from evolutionary effects.

The amplitude of the sinewave in the $O-C$ curve will depend almost entirely on the mass of the third body ($M_{\rm 3rd*}$) and its orbit as represented by the orbital period ($P_{\rm 3rd*}$) and its inclination ($i$).  The fitted $\dot{P}_{\rm 3rd*}$ will depend in the phasing of the interval $\Delta Y$, with the extreme values when the third body's conjunctions are near the middle of $\Delta Y$.  The fitted parabola to a part of a sinewave can be defined by the three points at conjunction and elongation.  With this, a limit on the maximum apparent period change is
\begin{equation}
|\dot{P}_{\rm 3rd*}| < 32 ~ \left( \frac{G M_{\rm total} P_{\rm 3rd*}^2}{4 \pi ^2} \right)^{\frac{1}{3}} ~ \frac{M_{\rm 3rd*}}{M_{\rm total}} ~ \frac{\sin(i)}{c}  ~ \frac{P}{P_{\rm 3rd*}^2}.
\end{equation}
Here, $M_{\rm total}$ is the sum of the masses of the three stellar components.  The inequality is to provide for the general case where the observing interval $\Delta Y$ is not centred on a conjunction.

In Table 9, I have tabulated $\dot{P}_{3rd*}$ for the stellar masses from Table 10, and for the extreme case with $M_{3rd*}$=1.0 M$_{\odot}$, $i$=90$\degr$, $P_{3rd*}$=2$\Delta Y$, and $\Delta Y$ is centred on the year of conjunction.  (For the cases of T Aur, V1500 Cyg, DQ Her, BT Mon, and U Sco, the limits from the $M_{V,q}$, including during the deep eclipses, limits any third body to have mass less than 0.5 M$_{\odot}$.)  These four extreme conditions are unlikely individually, and very unlikely to occur together in the same system, so any real third-body effect is likely greatly smaller than the tabulated values.  For the more typical expected cases, say with $M_{3rd*}$=0.1 M$_{\odot}$, $i$=30$\degr$, $P_{3rd*}$=10$\Delta Y$, the limit goes down by a factor of 330$\times$.

Now we are in a position to evaluate whether the observed $\dot{P}$ can be an artefact of a third body in orbit around the nova binary:  {\bf (1.)} The observed values of $|\dot{P}|$ are larger than the extreme limits for T CrB, T Pyx, V1017 Sgr, and U Sco.  {\bf (2.)} The observed $|\dot{P}|$ for HR Del, BT Mon, and RR Pic are nearly within a factor of two of the extremes of $\dot{P}_{\rm 3rd*}$, and this is very unlikely for even one system.  While it is possible for any one of these three novae to have the observed period change arising from a third-body orbit, the odds are strong that zero of the three has $\dot{P}$ dominated by such effects.  {\bf (3.)} DQ Her has the opposite problem, in that the measured $\dot{P}$ is so small that third-body effects are unlikely to be so small.  For example, $\dot{P}_{\rm 3rd*}$ can be made sufficiently small if $i$=0.05$\degr$, or if the third-body only has one-third the mass of Jupiter, or if the third body takes 25,000 years in its orbit.  Such is unlikely, but not impossible.  Whether or not DQ Her has a third body is irrelevant because its effects are measured to be negligibly small, and are effectively zero for all the comparisons of $\dot{P}$ with theory.  {\bf (4.)} The observed $\dot{P}$ for T CrB, T Pyx, and U Sco all changed substantially and effectively instantaneously at the time of their nova eruptions.  Such behaviour is impossible for the slow apparent period changes arising from a third-body.  In all, there is no real chance for third-body effects to be significant for 8 of the 14 novae with $\dot{P}$ measures.

For the remaining six novae (CI Aql, T Aur, QZ Aur, V1500 Cyg, V394 CrA, and IM Nor), it is possible that any one might have its $\dot{P}$ dominated by third-body effects.  Nevertheless, there are a variety of further arguments suggesting that few, if any, are so dominated.  CI Aql, QZ Aur, and V1500 Cyg have $\dot{P}$ within 7$\times$ of the extreme case $\dot{P}_{\rm 3rd*}$, so it is still unlikely for third body effects to get so large.  V394 CrA has a relatively poor measure of $\dot{P}$ with large error bars, so it is difficult to make a case for the dominance of any mechanism.  Five out of the remaining six novae have negative $\dot{P}$, which is understandable if some physical mechanism dominates, rather than the randomness of the third-body being near its superior conjunction 5-out-of-6 occasions.  And with this whole sample of 14 novae being selected independently of $\dot{P}$, then we would expect more than just DQ Her to have small $|\dot{P}|$ for a random distribution of orbital positions for the centre time of the intervals $\Delta Y$, unless few of the novae have substantial third-body effects in their $O-C$ curves.  Finally, the odds are low that any one of the six remaining novae happens to have a wide tertiary star with any sort of useable orbital period, phase, and inclination.  In all, I conclude that 8 of the 14 novae certainly have no significant third-body effects, while few (if any) of the remainder have significant third-body effects.

\subsection{Can $\dot{P}$ be an artefact of a stellar activity cycle?}

The `Applegate Mechanism' was devised in 1992 as a means to account for long-term period changes in close binaries by relating them to stellar activity cycles, with the resultant $O-C$ curve varying quasi-periodically like a sinewave (Applegate 1992).  This mechanism assumes that deep magnetic fields force shape changes in the companion star, with a changing quadrupole moment of the companion's mass resulting in small changes in $P$.

The Applegate Mechanism allows for an $O-C$ curve segment of duration $\Delta Y$ to appear to have a parabolic shape.  If the interval $\Delta Y$ is roughly centred on either the maximum or minimum of the stellar cycle, and if the cycle has a quasi-period of $>$2$\Delta Y$, then the observed part of the $O-C$ curve will appear approximately as a parabola.

{\bf (1.)} The nova companion stars are rather close to the main sequence (Knigge et al. 2011) with masses from 0.2 to 1.0 M$_{\odot}$ (Table 10).  (The exceptions are T CrB with its red giant companion, plus V1017 Sgr, U Sco, and V394 CrA with their subgiant companions.)  This corresponds to spectral classes of G, K, and early M stars.  Fortunately, in a massive observational campaign, the HK Project at Mount Wilson Observatory has measured the stellar cycle lengths for 20 main sequence stars from G2 to K7 and one G2 giant (Garg et al. 2019).  The main sequence stars have an average cycle length of 9.0 years, and a range from 4.6 to 14.9 years.  The giant star (as applicable for T CrB) has an average cycle length of 6.7 years.  These stellar activity cycles might be present in the $O-C$ curves as quasi-periodic oscillations with many full cycles, but only if the Applegate Mechanism has a significant amplitude.  That is, all the novae (other than U Sco) have $\Delta Y$ that are substantially longer than the longest known stellar cycle, so the activity cycle effects, if significant, must be visible as a sinewave with many cycles, rather than as a parabola.  For example, the parabolic shape of the 2001--2022 $O-C$ curve for CI Aql could only arise from stellar activity cycles if its cycle length is longer than 2$\Delta Y$, or $>$42 years, and such is effectively impossible.  The lack of any quasi-periodic sinewaves with periods under 14.9 years is proof that any stellar activity cycles have unmeasurably small effects in the $O-C$ curves.  With this, we can confidently eliminate any measurable stellar activity effects for all the novae (except U Sco).  

{\bf (2.)} The $O-C$ curves of T CrB, T Pyx, and U Sco show fast changes to $\dot{P}$ across their nova eruptions.  However, the sinewave effects of activity cycles cannot have fast changes in the $\dot{P}$.  Specifically, the stellar cycle is generated deep inside the companion star, where the nova effects cannot touch.  So, stellar activity cycles cannot account for the observed $\dot{P}$ values of T CrB, T Pyx, and U Sco.


{\bf (3.)} The Applegate Mechanism has long been sought in CVs.  An early and large search was made by Richman, Applegate, \& Patterson (1994), where they sought quasi-periodic changes in both period and magnitude for the 20 best-observed CVs.  No cyclic activity was identified with any useable confidence.  This proves that the stellar activity cycles are either rare, negligibly small, or both.  With the rarity of significant effects, it is improbable that any of my 14 nova will have their $\dot{P}$ values dominated by the Applegate Mechanism.

{\bf (4.)} Several serious problems indicate strong doubts about the existence of the Applegate Mechanism, at least at any significant or measurable level:  {\bf (4a.)} The phenomenon of quadrupole changes in the mass distribution in single stars or in binaries has never been observed.  Most critically, quadrupole changes in our own Sun, with its $\sim$11 year sunspot cycle, have never been measured, despite very exacting measures of the solar radius and helioseismology over multiple cycles.  If our Sun does not show the hypothetical effect to small degree, then it is difficult to think that any other main sequence star will show a large effect.  {\bf (4b.)} Quasi-sinusoidal variations in $\dot{P}$ for CVs have never been observed and confirmed.  That is, despite broad searches, no Applegate Mechanism has ever been confirmed, with this pointing to the non-existence of any significant effect.  {\bf (4c.)} The models  have the omission of any physics that produces the critical transfer of angular momentum between the inner and outer regions of the companion star.  The stated strength of the interior magnetic field is not taken from a physics calculation of stellar interiors, but rather is put in by hand so that effects might be visible.  We are never told the configuration of the magnetic field, nor how any such could provide a torque in the needed direction, nor how the magnetic field changes through the activity cycle, nor how the magnetic field can be as large as posited, nor where the energy comes from to spin up-and-down the outer envelope of the star.  The fundamental physics is missing from the one most critical part of the models.  The detailed models of the Applegate Mechanism (Applegate 1992: Richman et al. 1994; V\"{o}lschow et al. 2018) all input the size of the transfer of angular momentum as an un-evidenced assumption.  With this, the size and the existence of the Applegate Mechanism is dubious.

\subsection{Testing the predictions of the Magnetic Braking Model}

The MBM makes specific predictions as to the value of $\dot{P}_{MBM}$ as a unique function of $P$.  Knigge et al. (2011) states `Theoretically, all CVs with initially unevolved donors are expected to quickly join onto a unique evolution track, whose properties are determined solely by the mechanism for AML [angular momentum loss] from the system (Paczy\'{n}ski \& Sienkiewicz 1983; Ritter \& Kolb 1992; Kolb 1993; Stehle et al. 1996).'  The MBM predictions for their `optimal' model are given in Table 4 of Knigge et al., and plotted as the thick blue line in the upper-left of the lower panel in Figure 9.  These predictions can be directly tested with my measured $\dot{P}$.  Indeed, this is the first direct test of MBM predictions that I am aware of.


The MBM predictions fail by large factors for 10 of my 13 novae with useable $\dot{P}$ measures\footnote{V394 CrA has such a large uncertainty that it is not a useful test case.}:   {\bf (1.)} MBM requires $\dot{P}_{MBM}$ to always be negative\footnote{The exceptions are for period bounce systems and for a brief interval while passing over the edge of the Period Gap, with both cases certainly not being applicable to any of the 14 novae.}.  But HR Del, DQ Her, IM Nor, RR Pic, T Pyx, and V1017 Sgr all have positive $\dot{P}$.  {\bf (2.)} The observed $\dot{P}$ is greatly more negative than is possible from MBM for CI Aql, V1500 Cyg, U Sco, and T CrB.  These are significantly failed predictions for MBM.  {\bf (3.)} The novae T CrB, T Pyx, and U Sco all have large and significant changes in $\dot{P}$ that occur closely around the time of their nova events.  Such is not possible within the MBM.  All the failed predictions are summarized in Table 11.

Exceptions can be attempted for some of these counterexamples.  T Pyx is now in a weird special state (wherein its 1866 CN event initiated a century-long very-high accretion state resulting in RN events that are now tapering off) not imagined by the MBM of Knigge et al.  With a realistic measure for the T Pyx $\dot{M}$, its predicted $\dot{P}_{mt}$ can be raised to the level of agreement with observation.  The same argument for exception can be made for IM Nor.  That is, IM Nor also has a very high accretion rate\footnote{IM Nor has $\dot{M}$ close to the maximum, as shown by its very luminous $M_{V,q}$ (Patterson et al. 2022) and by being a recurrent nova.  Given the similarities between T Pyx and IM Nor, a reasonable suggestion is that IM Nor also suffered a normal CN eruption centuries ago that kicked off the current episode of RN events.  Such would provide an easy explanation for why IM Nor does not have the small $\dot{M}$ as required by the MBM for such a short $P$.}, so also has a high $\dot{P}_{mt}$ that dominates and is expected to produce a period change comparable to that observed.  So, despite the MBM prediction, the specific case of IM Nor does reasonably match the predictions of standard theory.  Another possible exception for consideration is DQ Her.  Within the idealized evolutionary scheme of the MBM, the predicted value for the period of DQ Her is $-$14.2 in units of 10$^{-13}$, in strong disagreement with the observed value of $+$1.3$\pm$1.2 in the same units.  But if we use standard theory for the specific case of DQ Her, the predicted period change is $+$62.1 in units of 10$^{-13}$, still in large and significant disagreement with the observed value.  Depending on whether we use the idealized evolutionary scheme of the MBM or the standard theory for the specific case of DQ Her, we predict period changes on opposite sides of the observed value.  Nevertheless,  we are left with the strong result that the predicted period change disagrees greatly and significantly from the observed $\dot{P}$.

The deep problems with MBM can be recognized by comparing the cases of the three `sister' novae HR Del, DQ Her, and T Aur.  These three have closely similar $P$ (all near 0.20 days), similar $M_{\rm WD}$ (all near 0.7 M$_{\odot}$), similar $M_{\rm comp}$ (all near 0.5 M$_{\odot}$), similar accretion rates (all near 1$\times$10$^{-8}$ M$_{\odot}$ yr$^{-1}$), and similar $R_{\rm comp}$ (all near 0.53 R$_{\odot}$).  The CN eruptions of all three have similar light curve classes (all from the low-$M_{\rm WD}$ classes of D and J), similar long $t_3$ decline rates (all from 84 to 231 days), the same spectral class (all Fe {\rm II}), similarly low absolute magnitude at peak (all near -6.5 mag), and similar dust formation (producing visible nova shells).  All the critical properties are close for these three systems.  MBM requires that they all should have the same $\dot{P}$.  Contrarily, HR Del has a large {\it positive} $\dot{P}$, while DQ Her has a {\it near-zero} value, and while T Aur has a large {\it negative} $\dot{P}$.  At best, MBM can only represent one-out-of-three of these sister-novae.  This illustrates the problem of MBM in a nutshell.

\subsection{Can averaging over time save the MBM?}

A possible explanation to save the MBM is to recognize that my $\dot{P}$ measures are from observing intervals of $\Delta Y$ up to 100 years in duration, while the MBM is only concerned with the {\it average} behaviour over time-scales of order 10-million years (Knigge et al. 2011, Fig. 5).  MBM would be saved if the widely scattered $\dot{P}$ measures happened to average out to the MBM predictions on sufficiently long time-scales.  I know of no mention of this possibility in the literature. 

The deviations between the observations ($\dot{P}$) and theory (either $\dot{P}_{MBM}$ or $\dot{P}_{total}$) range from large-and-negative to large-and-positive.  These observed deviations range over an additive effect from $-$10$^{-9}$ to $+$10$^{-9}$ for the common period range of novae (see Fig. 10).  This is to be compared to the MBM predicted range of roughly $-$10$^{-13}$ to $-$10$^{-11}$.  The deviations are $>$100$\times$ larger than the basic MBM effects.  So any mechanism that is creating the scatter must be symmetric (positive and negative), must have the typical size of the scattering effect be $>$100$\times$ larger than the theory effects, and must average close to zero in most nova systems.

How accurate or consistent must this time-average be?  A minimum criterion is that the nova-to-nova variation in the time-average must be sufficiently small so as to produce a fairly sharp-edged Period Gap.  Figure 14 of Knigge et al. (2011) shows that a 2$\times$ variation in the strength of the magnetic braking component will lead to an unacceptably poor sharpness of the upper edge of the Period Gap.  (To be specific, the edge varies from 2.8 to 3.6 hours as their scaling parameter varies from 0.25 to 1.0, so any time-averaging that has a residual variation of 2$\times$ or more will not produce the observed sharpness of the Period Gap edge.)  For CVs just above the Period Gap, the MBM gives a $\dot{P}$ of $-$5.1$\times$10$^{-13}$.  To keep the edge of the Period Gap sufficiently narrow, the $\dot{P}$ must be in the range $-$2.5$\times$10$^{-13}$ to $-$10$\times$10$^{-13}$.  The width of this range is 7.5$\times$10$^{-13}$, and any averaging must return a value with an accuracy better then this limit.

A severe problem arises for the time-averaging idea because the novae do not have enough time for averaging to get the required small scatter.  For my observations over $\Delta Y$, I see no changes in $\dot{P}$ over intervals up to 100 years, so novae appear to have episodes of relatively constant period changes lasting $>$100 years.  Presumably, each nova will have many episodes over the 10$^7$ year averaging interval, with these combining to produce a mean close to zero.  The number of episodes is $<$10$^5$.  Averaging over this many episodes will reduce the scatter of the mean by a factor of $\sqrt{10^5}$.  For a $\dot{P}$ scatter of 10$^{-9}$ for individual episodes, the $\langle \dot{P} \rangle$ will have a scatter of 10$^{-11.5}$.  The full range (i.e., two-sided) is 6.3$\times$10$^{-12}$.  This is to be compared to the maximum full range of 7.5$\times$10$^{-13}$ required to keep narrow the edge of the Period Gap.  Importantly, the best accuracy of averaging $\dot{P}$ over ten-million years can only narrow the range down to 8.3$\times$ larger than required.  Time-averaging cannot reduce the observed scatter of $\dot{P}$ by enough to explain the Period Gap.

This severe problem (how CVs average their period changes so as to produce a narrow-edged Period Gap) is broader than simply time-averaging some unknown mechanism, labelled as $\dot{P}_{\rm X}$, to have a spread of less than 2$\times$.  In general, to keep a narrow-edge, the scatter of $\langle \dot{P} \rangle$=$\langle \dot{P}_{\rm total} \rangle$+$\langle \dot{P}_{\rm X} \rangle$ must be kept small both as an average over time for individual systems plus as an average over all the individual CVs.  For an unknown mechanism, the standard theory cannot predict the scatter of $\langle \dot{P}_{\rm X} \rangle$.  For the $\langle \dot{P}_{\rm total} \rangle$ contribution, the individual CVs should have only modest variability over time, so averaging over ten-million years should return small scatter.  However, for standard theory, $\langle \dot{P}_{\rm total} \rangle$ will vary substantially from CV-to-CV, making for a substantial scatter in $\langle \dot{P} \rangle$.  The MBM solves this problem by insisting that for a given $P$ all CVs have the same stellar masses, the same accretion rates, the same $\Delta P$, and the same recurrence time-scales\footnote{The failure of these predictions is so extreme as to constitute a refutation of the MBM scheme of a single unique evolutionary track for CVs.  That is, for any given $P$, the observed properties vary over a large range, in defiance of the strong requirements of the MBM model.  For CVs just above the Period Gap, $M_{\rm WD}$ is observed to vary from 0.4 to 1.3 M$_{\odot}$, while $M_{\rm WD}$ is observed to vary over the range 0.20 to 0.73 M$_{\odot}$ (Ritter \& Kolb 2003; Pala et al. 2022).  This greatly violates the conditions and predictions of the MBM model.  There is no possibility to average out these masses over some hypothetical cycle.  Further, the measured $\dot{M}$ for CVs of the same $P$ varies with a range of around 1000$\times$.  For example, just above the Period Gap, the accretion rate has a range from 10$^{-10.5}$ to 10$^{-7.5}$ M$^{\odot}$ yr$^{-1}$ (Patterson 1984; Pala et al. 2022).  With this, we know that the basic MBM scheme of a single track for evolution is not working in practice, instead failing its predictions by up to three orders-of-magnitude.}.  The standard theory for $\dot{P}$ recognizes that CV properties can change greatly from CV-to-CV, and calculates the theoretical $\dot{P}_{\rm total}$ for each system.  The scatter in $\dot{P}_{\rm total}$ will necessarily be large due to the known CV-to-CV variations in just $M_{\rm WD}$ and $M_{\rm comp}$.  In particular, for a given range of $P$, say just above the Period Gap, the observed $M_{\rm WD}$ varies roughly uniformly between 0.4--1.3 M$_{\odot}$, while $M_{\rm comp}$ varies between 0.20--0.73 M$_{\odot}$ (Ritter \& Kolb 2003; Pala et al. 2022).  For CVs just above the Period Gap, those with well-measured $P$, $M_{\rm WD}$, and $M_{\rm comp}$ from Ritter \& Kolb, I calculate $\dot{P}_{\rm total}$ assuming that all the other CV properties (including $\dot{M}$, $M_{\rm ejecta}$, and $\tau_{\rm rec}$) are held constant for all the systems.  With these calculated $\dot{P}_{\rm total}$ values, the total range is 4$\times$.  That is, just the known variations in the stellar masses makes for a variation of 2$\times$ around the $\langle \dot{P} \rangle$ value.  This by itself is enough to smear the edge of the Period Gap to wider than is observed.  And this smearing will not be changed by averaging over time, since the masses do not change greatly over each averaging interval.  The additional scatter of $\dot{P}_{\rm total}$ caused by the large observed variations in $\dot{M}$ (and hence variations in $\dot{P}_{\rm ml}$) are greatly larger still.  For CVs just above the Period Gap, the observed $\dot{M}$ varies from under 10$^{-10.5}$ to 10$^{-7.5}$ M$_{\odot}$ yr$^{-1}$ (Patterson 1984; Pala et al. 2022), so the $\dot{P}_{\rm mt}$ varies by up to a factor of 1000$\times$, and $\dot{P}_{total}$ varies from $-$1$\times$10$^{-12}$ to $+$2$\times$10$^{-10}$ for real CVs with the same $P$.  So just ordinary observed CV-to-CV variations of properties makes for a large scatter of the predicted $\dot{P}_{\rm total}$, with this being more than large enough to make for the predicted edge of the Period Gap to be significantly wider than is observed.  In all, the averaging of the standard theory $\dot{P}_{total}$ cannot reduce the scatter enough so that $\langle \dot{P} \rangle$ can follow any single evolutionary track so as to produce a narrow edge for the Period Gap.  This can only be viewed as a critical failure of the standard theory, where the key prediction of a Period Gap is not actually realized from the actual population of CVs.

\subsection{A new mechanism is required to explain the large $\dot{P}$ scatter} 


The extreme scatter about the theory predictions forces us to realize that some unmodelled mechanism is at work for most novae.  That is, some additional mechanism (past the first three in Section 11.1) must be introducing this scatter.  The large deviations from the MBM predictions must arise from some physical mechanism, variations on the known physical mechanisms cannot explain the deviations, so the physical mechanism for the deviations is not one of the modelled mechanisms.  The forced conclusion is that there must be some now-unidentified physical mechanism that is not included in MBM or standard theory.  This strong conclusion has far-reaching implications.

A critical implication is that the MBM and standard theory is incomplete, with the missing mechanism dominating by orders of magnitude over the modelled mechanisms.   A realist would acknowledge that the MBM calculations are often off by large factors, but expect that the incomplete standard theory has captured the basic situation correctly.  Realists in our community should be actively working on the theory of new physical mechanisms, and new evolutionary models should be constructed with the empirical reality of the observed period changes.  The primary question in CV studies is their evolution, and this is driven by the period changes, now revealed to be missing the dominant effects.  Thus, the measured $\Delta P$ and $\dot{P}$ should be setting the top of the agenda for future work on CVs.


\section{ACKNOWLEDGEMENTS}

I thank Arto Oksanen, Gordon Myers, Shawn Dvorak, Franz-Josef Hambsch, David Cejudo Fernandez, Tonny Vanmunster, Bill Goff, James Boardman, Geoffrey Stone, Libert Monard, Peter Nelson, Richard Sabo, Joseph Ulowetz, Walt Cooney, Gary Walker, and Lew Cook for their photometry that produced many minimum timings.

I thank Tom Maccarone, Christian Knigge, Saul Rappaport, and Juhan Frank for valuable, long, and detailed discussions.

\section{DATA AVAILABILITY}

All data are publicly available, either in the cited literature, the cited databases, or in the tables.


{}

\bsp	
\label{lastpage}
\end{document}